\documentclass[aps,prb,twocolumn,a4paper,twoside,showkeys,showpacs,floatfix,final]{revtex4}
\usepackage{amsmath,amssymb,amsfonts}
\usepackage{graphicx}
\usepackage[english]{babel}
\usepackage{color}
\usepackage{tabularx}
\usepackage{txfonts}

\newcommand{\Sb}{\mathbf{S}} \newcommand{\Jp}{$J-J^\prime$}
\newcommand{\etal}{\emph{et al.}}
\def\gsim{\mathrel{\rlap{\lower4pt\hbox{\hskip1pt$\sim$}}\raise1pt\hbox{$>$}}}

\begin{document}
\title{Comprehensive quantum Monte Carlo study of the quantum critical
  points in planar dimerized/quadrumerized Heisenberg models}
\author{Sandro Wenzel}\email{wenzel@itp.uni-leipzig.de}
\author{Wolfhard Janke} \email{janke@itp.uni-leipzig.de}
\affiliation{Institut f\"ur Theoretische Physik and Centre for
  Theoretical Sciences (NTZ), Universit\"at Leipzig, Postfach 100 920,
  D-04009 Leipzig, Germany}

\date{\today}

\begin{abstract}
  We study two planar square lattice Heisenberg models with explicit
  dimerization or quadrumerization of the couplings in the form of
  ladder and plaquette arrangements. We investigate the quantum
  critical points of those models by means of (stochastic series
  expansion) quantum Monte Carlo simulations as a function of the
  coupling ratio $\alpha=J^\prime/J$. The critical point of the
  order-disorder quantum phase transition in the ladder model is
  determined as $\alpha_\mathrm{c}=1.9096(2)$ improving on previous
  studies. For the plaquette model, we obtain
  $\alpha_\mathrm{c}=1.8230(2)$ establishing a first benchmark for
  this model from quantum Monte Carlo simulations. Based on those
  values, we give further convincing evidence that the models are in
  the three-dimensional (3D) classical Heisenberg universality class.
  The results of this contribution shall be useful as references for
  future investigations on planar Heisenberg models such as concerning
  the influence of non-magnetic impurities at the quantum critical
  point.
\end{abstract}


\maketitle

\section{Introduction}

The study of quantum effects in magnetism is an ongoing and
fascinating part of physics
research.\cite{quantummagnetism-Richter,sachdev-2008-4} Within this
area, the low-dimensional $S=1/2$ Heisenberg antiferromagnet plays an
eminent role. This is partly because it correctly describes aspects of
cuprate superconductors and is thus implemented in nature.  Second,
the Heisenberg model and variations have seen a lot of investigations
as toy models where quantum fluctuations lead to unexpectedly rich and
exotic ground-states (such as valence bond solids and valence bond liquids). Recent experiments in optical lattices\cite{Trotzky:2008a}
further provide the perspective to directly implement those models in
a pure environment thereby enabling a direct experimental access and
comparison between theory and measurements.

In two-dimensional (2D) Heisenberg models, the Mermin-Wagner theorem
forbids phase transitions to occur at $T\neq 0$, yet quantum
fluctuations may lead to a transition between ground states, for
example from an ordered N\'eel to a disordered state at zero
temperature. Such transitions are termed quantum phase
transitions.\cite{sachdev:qpt,vojta-2003-66} One way in which quantum
fluctuations can destroy order is for example provided via frustration
of bonds (next-nearest-neighbor couplings) or the inclusion of 4-site
interactions.\cite{sandvik-2007}

In a second mechanism, competition between locally varying
nearest-neighbor bonds of the same kind has been identified to cause
quantum phase transitions, for example by favoring the formation of
spin singlets. An important class of models in which the latter
mechanism is at work are the so called dimerized Heisenberg models
(where we also use the term for extended models with quadrumerization,
etc.), where the competition among couplings is explicitly introduced
in a geometric manner.  Apart from their relevance as simple models
for quantum phase transitions such systems have been in recent focus
in connection with Bose-Einstein condensation of
magnons.\cite{giamarchi-2008-4} A prominent example of dimerized
models is the $S=1/2$ bilayer Heisenberg system
\cite{PhysRevLett.72.2777,PhysRevB.51.16483,PhysRevLett.81.5418,PhysRevB.61.3475,collins:054419}
which consists of two $L\times L$ layers, where the inter-layer
coupling $J_\perp$ can be different from the intra-layer coupling $J$
(both couplings antiferromagnetic). Competition between $J_
\perp$ and $J$ can drive a quantum phase transition.

Due to progress and availability of unbiased and efficient methodological
schemes\cite{evertz-2003-52} some numerical contributions in the literature were
lately pushing results on those bilayer systems to  unprecedented
accuracy for quantum models, allowing for very detailed studies in the
quantum critical regime. Following the high precision
study on two bilayer systems by Wang \etal,\cite{wang:014431}
H\"oglund and Sandvik could for example report on anomalous
response\cite{hoeglund:027205} of non-magnetic impurities, for which
an accurate knowledge of the quantum critical point was a
prerequisite. The overall interest on such impurity based questions is
growing,\cite{sachdev-1999-286,anfuso-2006-96,sushkov_impurity}
therefore asking for the general availability of more detailed data
also in other systems.

While the level of accuracy has reached a very high quality for
bilayer systems, this is not equally the case for planar geometries.
After the seminal simulation of the CaVO lattice by Troyer \etal{},\cite{troyer-1997-66} only the coupled ladder model was
considered in more detail\cite{PhysRevB.65.014407} using quantum Monte
Carlo (QMC) studies. A main result of these investigations was the
confirmation of the critical exponents predicted by field
theory.\cite{ChakravartyPRL,chubukov_prb}

In an effort to systematically improve and extend these results to
other planar Heisenberg models, we have recently started with a
contribution\cite{wenzel-JJp-2008} reporting on peculiar and
non-universal features of a particular dimerized model, called the
\Jp{} or staggered
model.\cite{PhysRevB.53.2633,PhysRevLett.61.2484,PhysRevB.61.14607,tomczak,darradi-2004-16}
In Ref.~\onlinecite{wenzel-JJp-2008}, our presentation is based on a
detailed scaling analysis at criticality and a comparison between several
dimerized models including bilayer and planar geometries. As a
prerequisite to this comparison, we have also presented new but
preliminary results on the ladder and plaquette Heisenberg model
without showing any details of our numerical data nor its data
analysis. An in-depth study of these models on its own is, however,
useful for several reasons.  Apart from the aforementioned
motivation concerning impurities, new benchmark results shall be useful
for thermodynamic considerations in the quantum critical regime and
for further developing and testing novel algorithms and numerical
techniques.

\begin{figure}[t]
\centering
\begin{minipage}{0.55\columnwidth}
\includegraphics[width=\textwidth]{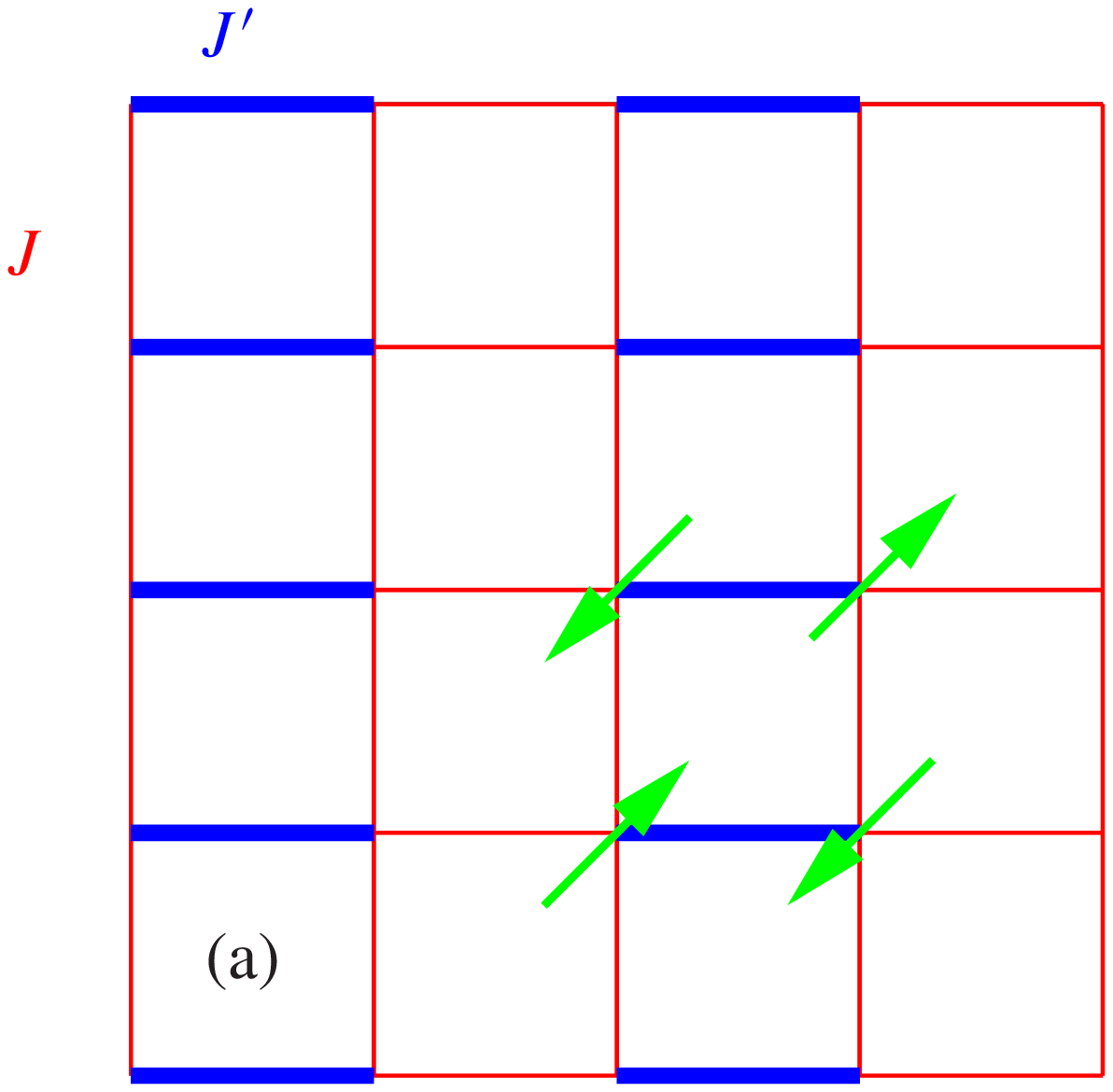}
\end{minipage}\vspace{0.4cm}
\begin{minipage}{0.55\columnwidth}
\includegraphics[width=\textwidth]{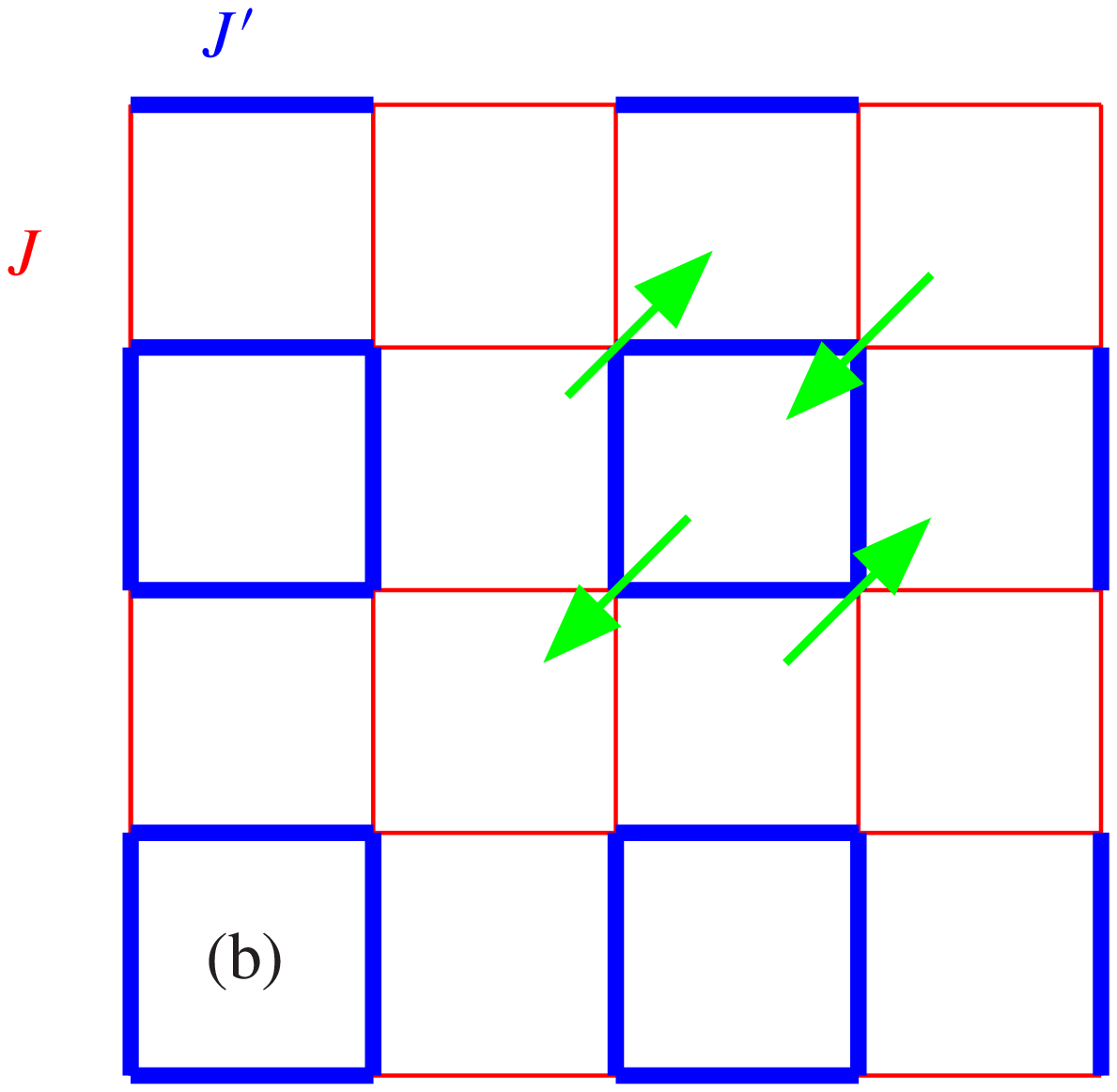}
\end{minipage}
\caption{\label{fig:dimers}(Color online) (a) Visualization of the
  ladder model on the two-dimensional square lattice. The quantum spin
  ($S=1/2$) degrees of freedom live on a square lattice with different
  nearest-neighbour couplings $J$ and $J^\prime$ (thin and thick
  bonds). The lattice bonds corresponding to couplings $J^\prime$ are
  denoted as $\langle i,j\rangle^\prime$ in the Hamiltonian. (b)
  Similar for the plaquette model, favoring quadrumer formation.
Both systems are studied using periodic boundary conditions.}
\end{figure}

In order to close this existing gap, we consider in this paper
the critical points of the ladder and plaquette models defined by the
Hamiltonian
\begin{equation}
\label{egn:hamiltonian}
\mathcal{H}=J\sum_{\langle i,j\rangle}\Sb_i \Sb_j+J^\prime \sum_{\langle i,j\rangle^\prime}\Sb_i \Sb_j\,.
\end{equation}
Here, $\Sb_i=(1/2)\,(\sigma_x,\sigma_y,\sigma_z)$ denotes the usual
spin-$1/2$ operator at lattice site $i$, and $J$ and $J^\prime$ the
antiferromagnetic coupling constants defined on bonds $\langle i,j
\rangle$ and $\langle i,j \rangle^\prime$, respectively. The
arrangements of the bonds on the square lattice of size $L$ in both
directions can be seen in Fig.~\ref{fig:dimers}.
Let us define the quantity $\alpha=J^\prime/J$ as the ratio of the two
competing couplings. For $\alpha>\alpha_\mathrm{c}>1$
the systems will be disordered and gaped due to formation of
spin-singlets. For $\alpha_l<\alpha<\alpha_\mathrm{c}$ the
systems possess N\'{e}el order and there is no gap. Here $\alpha_l$ is some lower boundary at which a second transition can take place. In this regard, it is interesting to note that long range N\'eel order even for $\alpha=1$ was only recently established rigorously.\cite{loewNeel}
With $\alpha_\mathrm{c}$ we denote the quantum critical point.
Throughout this work we fix $J=1$ and study the transition from the N\'eel
to a disordered state, when $\alpha$ (or $J^\prime$) is
increased.\footnote{For the plaquette model, there is a symmetric
  transition, when $\alpha$ is decreased. For the ladder model this
  second transition is between a ladder structure and N\'{e}el order.}

Let us first summarize some previous work on the subject.  Early
contributions on the ladder model were done by Singh \etal{},\cite{PhysRevLett.61.2484} who used
series expansions to access the critical
point. Numerically oriented work followed
from Katoh and Imada\cite{katoh1993} and was later improved by
Matsumoto \etal{}\cite{PhysRevB.65.014407} in a detailed QMC study which had its major objective in studying the $S=1$ case. For $S=1/2$, to
our knowledge the best known value for the critical coupling is taken
from that paper as $\alpha_\mathrm{c}=1.9107(2)$ (which is the inverse
of $0.52337(3)$), together with an estimate of the critical exponent
$\nu=0.71(3)$. The latter result is often used/quoted in favor of $O(3)$
universality  based on field theory. The
$S=1/2$ ladder model has been further investigated in three dimensions
(3D) in connection with field induced phenomena and Bose condensation
of magnons.\cite{nohadani-2005-72,nohadani-2004-69} The effects of
random site dilution in the dimerized phase were also
studied.\cite{chitoshi_dilution}
Quite generally, the coupled ladder model is nowadays often used as a
paradigmatic model in discussions of quantum phase transitions and
quantum magnetism.\cite{sachdev-2008-4,senthil-2005-74}

Less is known about the plaquette model, which was studied before
mainly analytically or with series expansions.\cite{koga,sirker:134409,singh-1999-60} A recent study on
the quadrumerized Shastry-Sutherland-model\cite{wessel:shastry}, using mainly
exact diagonalization methods, also contains a (hidden) QMC
estimate of the critical coupling $\alpha_c\approx 1.82$ for the pure
plaquette model. Additionally, the plaquettized model returned into
focus using a numerical scheme called contractor renormalization
(CORE) method.\cite{capponi:104424} Still, it lacks a detailed quantum
Monte Carlo investigation as presented in this paper.

The reason to reconsider the ladder model is threefold. First, we like
to test our algorithm and approach on known models. Our second
motivation is to complement the description of the phase transition in
the ladder model beyond to what was done earlier. This includes the
extension to different critical quantities, inclusion of corrections
in the finite-size scaling analysis and calculation of critical
exponents not considered before. Our aim is also to make the value of
$\nu$ more accurate for definite interpretation in favor of $O(3)$
universality. Lastly, a major objective is to derive results which we
partly presented in Ref.~\onlinecite{wenzel-JJp-2008}, as the
dimerized ladder model is so similar to the staggered model.

We organize our paper as follows. In Sec.~\ref{sec:methods} we shortly
present our implementation of the QMC method and data-analysis
approaches. Standard observables used to detect the critical point are
defined and discussed. A detailed presentation of our numerical data
with a focus on the critical point is given in
Sec.~\ref{sec:analysis}. Section~\ref{sec:scaling} contains a
finite-size scaling analysis of the critical exponents and a summary
is given in Sec.~\ref{sec:conclusions}.

\section{\label{sec:methods} Simulation methods and finite- size scaling}
\subsection{Quantum Monte Carlo simulations}

In this work, we report on simulations based on our implementation of
the stochastic series expansion (SSE) method by Sandvik and
Kurkij\"arvi.\cite{PhysRevB.43.5950} Due to its discrete nature, this
QMC scheme is a convenient and powerful method to
implement. The central idea of SSE is to sample the series expansion
of the partition function
\begin{equation}
\label{eq:z}
Z=\mathrm{tr}\left[\exp(-\beta \mathcal{H})\right]=\sum_\alpha \sum_n \frac{(-\beta)^n \langle \alpha| \mathcal{H}^n |\alpha\rangle}{n!}\,,
\end{equation}
with $n$ being the expansion order, $|\alpha\rangle$ a basis state of
the spin space, and $\beta$ the inverse temperature. While the
original algorithm used local Metropolis-type updates, major
improvements were achieved by introducing cluster or operator loop
updates.\cite{PhysRevB.59.R14157} Our own implementation is based on
the directed loop\cite{PhysRevE.66.046701} generalization together
with additional ideas described by Alet \etal.\cite{alet:036706} The
recent incorporation of the Wang-Landau
method\cite{PhysRevLett.86.2050} into the SSE
scheme\cite{PhysRevLett.90.120201} allows the use of multihistogram
techniques on QMC data \cite{ferrenberg:multi,Troyer_multihist} which
is useful to obtain unbiased continuous curves through data points, a
feature which we use partly in our data analysis.

\begin{figure}[t]
\centering
\includegraphics[width=0.9\columnwidth]{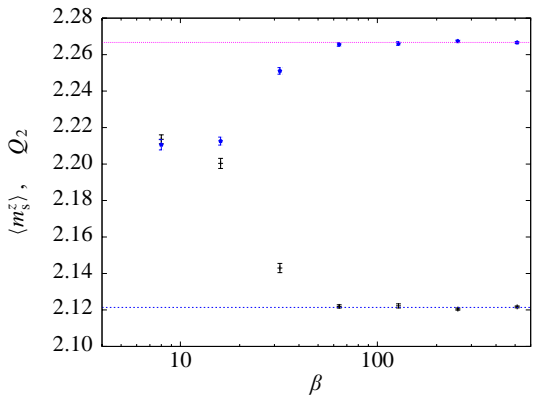}
\caption{\label{fig:conv}(Color online) Convergence test for the
  plaquette model at a system size $L=32$ and coupling $\alpha=1.82$
  displaying the two quantities $Q_2$ (upper curve) and $\langle |m_s^z| \rangle$ (lower curve). Ground-state
  properties are sampled for $\beta \gsim 100$. The staggered
  magnetization was multiplied by $25$ for convenience.}
\end{figure}

In order to access zero temperature properties of the spin system, all
simulations must be performed at sufficiently large $\beta$ so that
quantities of interest assume their ground-state value. In this
contribution this is done in a two stage procedure. For a chosen
lattice size, we check explicit convergence of observables by a $\beta$
doubling approach, i.e., we double $\beta$ until quantities agree
within error bars. Once a suitable $\beta$ is fixed for the chosen
size, standard aspect ratio scaling is employed.  Hence, we fix $\beta$
at lattice size $L$ according to $\beta_L=s L$, with $s$ being the
scale determined in the doubling scheme.\footnote{A slightly better
  way would be to test convergence for the largest lattice size and to
  simulate all smaller systems using the same temperature.}
Figure~\ref{fig:conv} shows a particular convergence test for a medium
sized lattice ($L=32$) indicating ground-state convergence for
$\beta\gsim 100$ for two exemplary observables defined below. This concrete test
was performed close to the critical point for the plaquette model
using $4\times 10^5$ sweeps. The inverse temperatures used in this
study are therefore rather large compared to some earlier studies.

\subsection{\label{sec:observables}Observables}

In order to determine the quantum critical point, we look at well-known
observables. Next to trivial quantities such as the average energy per
site $e$, we consider the staggered magnetization defined by
\begin{equation}
m^z_\mathrm{s}=\frac{1}{N}\sum_i^N S_i^z (-1)^{x_i+y_i}\,,
\end{equation}
where the sum runs over all $N=L^2$ lattice sites, together with the usual
Binder parameters
\begin{align}
Q_1 &=\frac{\langle (m_\mathrm{s}^z)^2 \rangle}{\langle |m_\mathrm{s}^z| \rangle^2}\,,\\
Q_2 &=\frac{\langle (m_\mathrm{s}^z)^4 \rangle}{\langle (m_\mathrm{s}^z)^2 \rangle^2}\,.
\end{align}
These parameters are dimensionless and they possess the property to
cross at the quantum critical point.
%Arguments that they take on
%universal values within a universality class should be taken with care
%since lattice structure, boundary conditions and the structure of the
%Hamiltonian may lead to non-universal features. The question whether
%$Q_2$ is universal at least for the periodic boundary conditions and
%the square lattice of our models will be discussed in the
%Sec.~\ref{sec:scaling}. 
Note that the staggered magnetization and the Binder parameters can be
determined quite efficiently by averaging over spin representations in
the operator direction\cite{PhysRevE.66.046701} of the SSE
representation. The brackets $\langle\dots\rangle$ therefore signify
$\langle m_\mathrm{s}^z \rangle \equiv \langle \langle m_\mathrm{s}^z
\rangle_\mathrm{op} \rangle_\mathrm{conf}$.

Second, we study the correlation length $\xi$ of the system. We employ
the standard second-moment approach, which uses the structure factors
$S(\mathbf{q})$ defined by
\begin{equation}
  S(\mathbf{q})=\frac{1}{N}\sum_{i,j} \exp(-\mathrm{i}\mathbf{q}(\mathbf{r}_i-\mathbf{r}_j))\langle S^z_i S^z_j \rangle\,,
\end{equation}
with $\mathbf{q}$ being a wave vector in Fourier space and
$\mathbf{r}_i$ the vector pointing to site $i$ on the real space
lattice. This quantity can be efficiently obtained for arbitrary
$\mathbf{q}$ during the diagonal update, as
\begin{equation}
  S(\mathbf{q})=\left\langle \frac{1}{Nn} \left(  \sum_{p}^{n-1} m_q[p] m_q[p]^\star \right)\right\rangle\,,
\end{equation}
where the index $p$ is running over the operator sequence having $n$
non-unit operators. The quantities $m_q[p]$ are defined at SSE operator
slice $p$ as $m_q[p] = \sum_i^N S^z_i[p] (\cos(\mathbf{q}
\mathbf{r}_i) - \mathrm{i}\sin(\mathbf{q}\mathbf{r}_i)) $ and
$m_q[p]^\star$ denotes its complex conjugate. The correlation length
is then estimated by%\footnote{Usually the prefactor is  $1/(2\sin(\pi/L))$.}
\begin{equation}
  \xi_y = \frac{L_y}{2\pi}\sqrt{\frac{S(\pi,\pi)}{S(\pi,\pi+2\pi/L_y)}-1}\,.
\end{equation}
For the anisotropic ladder model we expect $\xi_x \neq \xi_y$ on the
square lattice. We found it most useful to look at the correlation
length in $y$-direction of the system. This choice is arbitrary but
somehow motivated from Ref.~\onlinecite{wenzel-JJp-2008} because
$\xi_y$ showed good scaling for the staggered Heisenberg model. From
standard finite-size scaling theory we expect the quantity $\xi_y/L$
to cross for different lattice sizes at the quantum critical point. In
case of the symmetric plaquette model, an improved estimate for the
spatial correlation length can be obtained by taking
\begin{equation}
  \xi=\frac{1}{2}\left(\xi_x + \xi_y\right).
\end{equation}

Lastly, we consider the spin stiffness $\rho_s$ given
by\cite{PhysRevB.56.11678}
\begin{equation}
\rho_s=\frac{3}{4\beta}\langle w_x^2 + w_y^2\rangle\,,
\end{equation}
with $w_x$,$w_y$ being winding numbers defined by
\begin{equation}
  w_\lambda = (N^+_\lambda - N^-_\lambda )/ L_\lambda \quad (\lambda = x,y)\,.
\end{equation}
The symbols $N^+_\lambda$ and $N^-_\lambda$ represent the number of
operators of type $S^+_iS^-_j$ and $S^-_iS^+_j$ along the
$\lambda$-direction in the SSE configuration. The spin stiffness
measures the response in free energy upon a boundary twist on the staggered magnetization (the order parameter field $\theta$) and is also called superfluid density
in other contexts. At a quantum critical point in a 2D system it is
expected to scale as $\rho_\mathrm{s}\sim L^{d-2-z}$, where $z$ is the
dynamical critical exponent.\cite{PhysRevB.40.546,wang:014431}

\subsection{Finite-size scaling}
In this paper, we employ a variety of finite-size scaling methods to
determine various critical quantities from the quantum critical point
to the critical exponents. To this end, we make use of the standard
scaling ansatz in the vicinity of the critical point
\begin{equation}
  \label{eqn:scaling1}
  \mathcal{O}_L(t)=L^{\lambda/\nu}g_\mathcal{O} (tL^{1/\nu})\,,
\end{equation}
where $\nu$ is the critical exponent of the correlation length,
$\lambda$ the critical exponent of the quantity $\mathcal{O}$,
$g_\mathcal{O}(x)$ the scaling function, $t$ the reduced coupling
defined by $t=(\alpha-\alpha_\mathrm{c})/\alpha_\mathrm{c}$, and $L$
the lattice size.

Analysis to \eqref{eqn:scaling1} was performed in the previous QMC study on the ladder model in
Ref.~\onlinecite{PhysRevB.65.014407}. Here, we would like to go one step further and take leading
corrections to scaling into account. Apart from higher order terms
$O(1/L^2)$, the renormalization group (RG) then predicts a scaling of
the form
\begin{equation}
\label{eqn:scaling2}
  \mathcal{O}_L(t)=L^{\lambda/\nu}\left[ g_\mathcal{O} (tL^{1/\nu}) + L^{-\omega} g_\omega (tL^{1/\nu})\right]\,,
\end{equation}
where $\omega$ is the leading correction exponent and $g_\omega(x)$
another scaling function. Writing $g_\omega(x)=c(x) g_\mathcal{O}(x)$,
this becomes
\begin{equation}
\label{eqn:scaling3}
\mathcal{O}_L(t)=L^{\lambda/\nu}\left( 1 + c(x) L^{-\omega} \right) g_\mathcal{O}(x)\,,
\end{equation}
with $x=tL^{1/\nu}$ and a coefficient $c(x)$ depending on $x$. To
zeroth order, and for $x$ small we may set $c(x)\approx
c=\mathrm{const}$ and arrive at the usually employed form
\begin{equation}
\label{eqn:scaling4}
\mathcal{O}_L(t)=L^{\lambda/\nu}(1 + c L^{-\omega}) g_\mathcal{O}(x)\,.
\end{equation}
We consider \eqref{eqn:scaling4} as our primary ansatz in the data
analysis.

Note, that in the literature, another ansatz in form of 
\begin{equation}
\label{eqn:scaling5}
\mathcal{O}_L(t)=L^{\lambda/\nu}(1 + cL^{-\omega})g_\mathcal{O} (tL^{1/\nu} + dL^{-\phi/\nu})\,,
\end{equation}
has been discussed which represents an effective approximation to
\eqref{eqn:scaling2} in the vicinity of the quantum critical
point.\cite{beach-2005,wang:014431} Here, $\omega$ and $\phi$
represent \emph{effective} corrections, approximating the correct RG
behavior. In Ref.~\onlinecite{wang:014431}, which is closely related to
the present paper, the authors employed \eqref{eqn:scaling5} and
obtained results in excellent agreement with the expectations.  Here, we
will primarily employ \eqref{eqn:scaling4} and in some instances
compare our result to \eqref{eqn:scaling5}. In any case, we use this
procedure mainly to obtain the critical coupling
$\alpha_\mathrm{c}$.\footnote{For critical couplings, the method of
  Ref.~\onlinecite{beach-2005} produced reliable estimates which are
  consistent with other results in the literature.} We emphasize that
final results of critical exponents will be given as obtained from
ordinary scaling methods at the critical point ($x=0$), which are
described in Sec.~\ref{sec:scaling}.

Data analysis according to Eq.~\eqref{eqn:scaling4} is known as ``data
collapsing''. In practice, this can often be achieved by Taylor
expanding the scaling function $g_\mathcal{O}(x)$ for $x\to0$ into a
polynomial of the form
\begin{equation}
\label{eqn:expansion}
g_\mathcal{O}(x)=g_0 + g_1 x + g_2 x^2 + \dots\,.
\end{equation}
Using this ansatz, relation \eqref{eqn:scaling1} is turned into
\begin{equation}
\mathcal{O}_L(t)=L^{\lambda/\nu} \left(g_0 + L^{1/\nu} g_1 t + L^{2/\nu}  g_2 t^2 + \dots\right)\,,
\end{equation}
where all free parameters can then be determined by a nonlinear-fit of
the measured data. The generalization to \eqref{eqn:scaling4} is
obvious.

We have recently implemented a related method, which does not need to
make use of Taylor expanding the function
$g_\mathcal{O}(x)$.\cite{wenzel-2007} Using multihistogram techniques,
it is possible to directly perform a collapse of the data by
minimizing the weight function
\begin{equation}
\label{eqn:collapsequality}
  \sigma_\mathcal{O}^2=\int_{x_\mathrm{min}}^{x_\mathrm{max}} \mathrm{d}x \left[\overline{\widehat{\mathcal{O}}_L^2}(x) - \overline{\widehat{\mathcal{O}}}_L^2(x)\right]\,, 
\end{equation}
where
$\widehat{\mathcal{O}_L}(x)=\mathcal{O}_L(t)/(L^{\lambda/\nu}(1+cL^{-\omega}))$
and $x=tL^{1/\nu}$. With
$\overline{\widehat{\mathcal{O}}_L}(x)\equiv \sum_L
\widehat{\mathcal{O}}_L(x)/n_L$, we denote the average over $n_L$
lattice sizes.  For the quantities $Q_1$, $Q_2$, $\xi_y/L$, $\xi/L$,
and $\rho_\mathrm{s}L$ we have $\lambda/\nu=0$.
\begin{figure*}
\begin{minipage}{0.4\textwidth}
\begin{minipage}{0.9\textwidth}
\includegraphics[width=\textwidth]{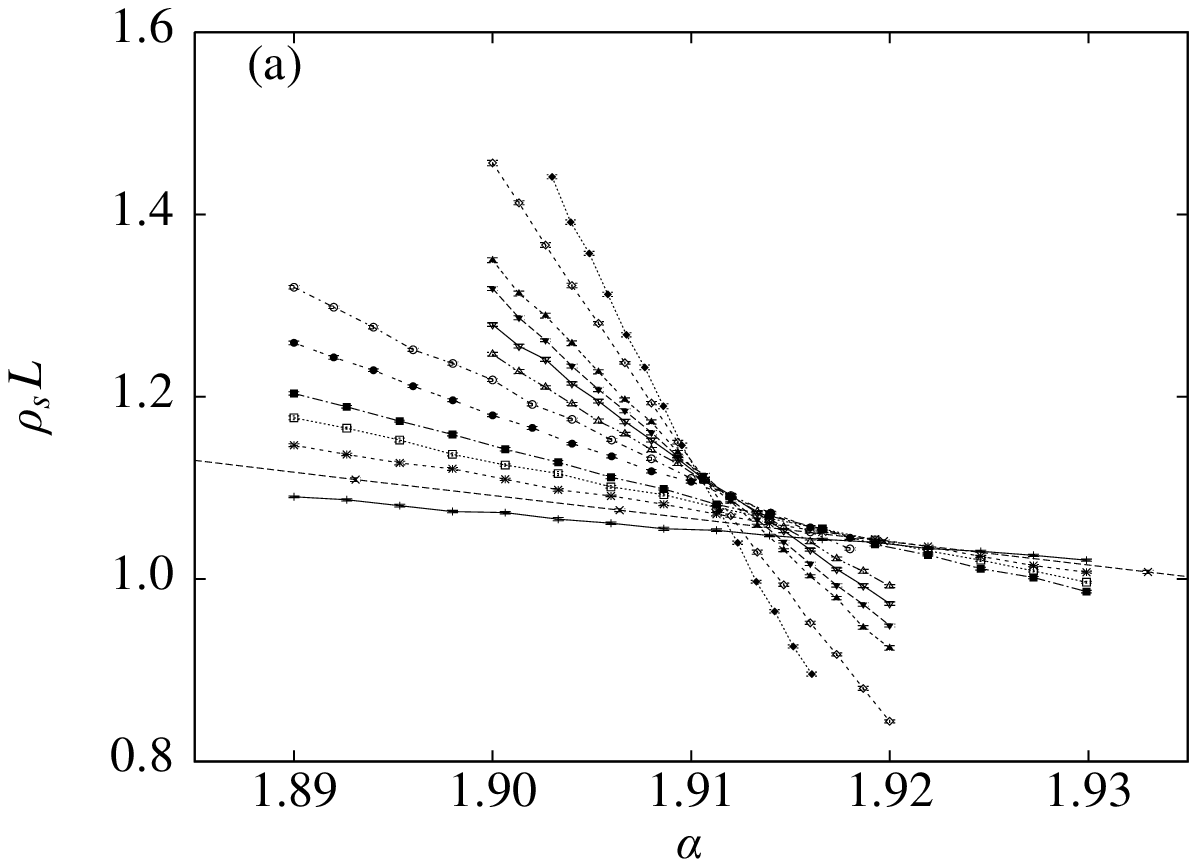}
\end{minipage}
\begin{minipage}{0.9\textwidth}
\includegraphics[width=\textwidth]{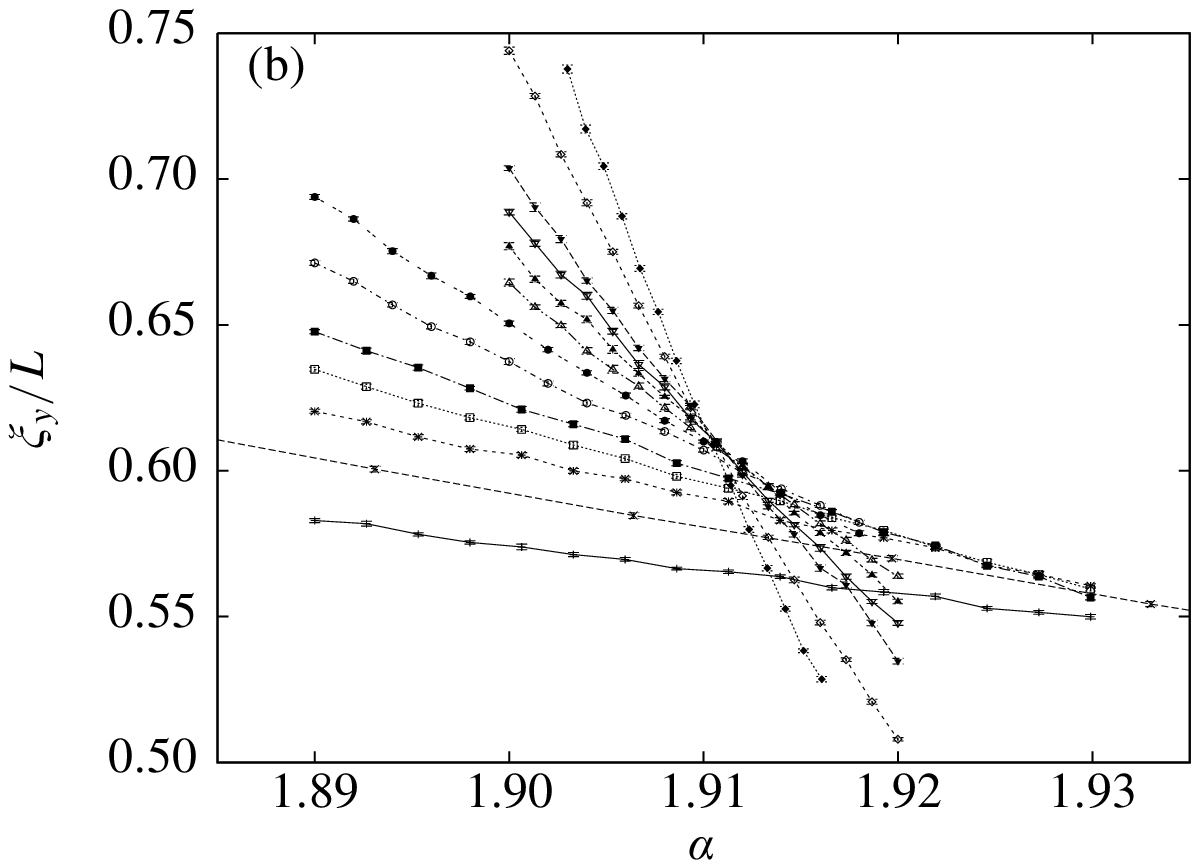}
\end{minipage}
\begin{minipage}{0.9\textwidth}
\includegraphics[width=\textwidth]{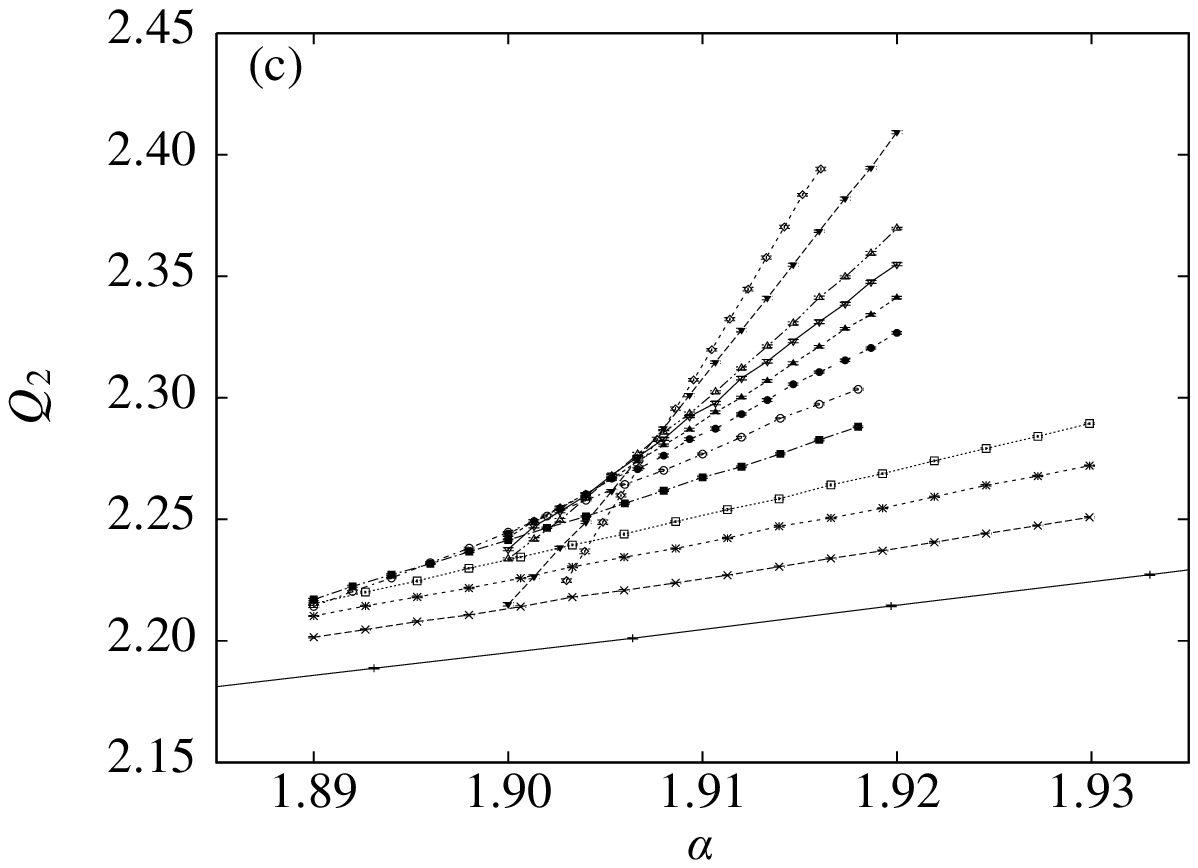}
\end{minipage}
\end{minipage}
\begin{minipage}{0.4\textwidth}
\begin{minipage}{0.9\textwidth}
\includegraphics[width=\textwidth]{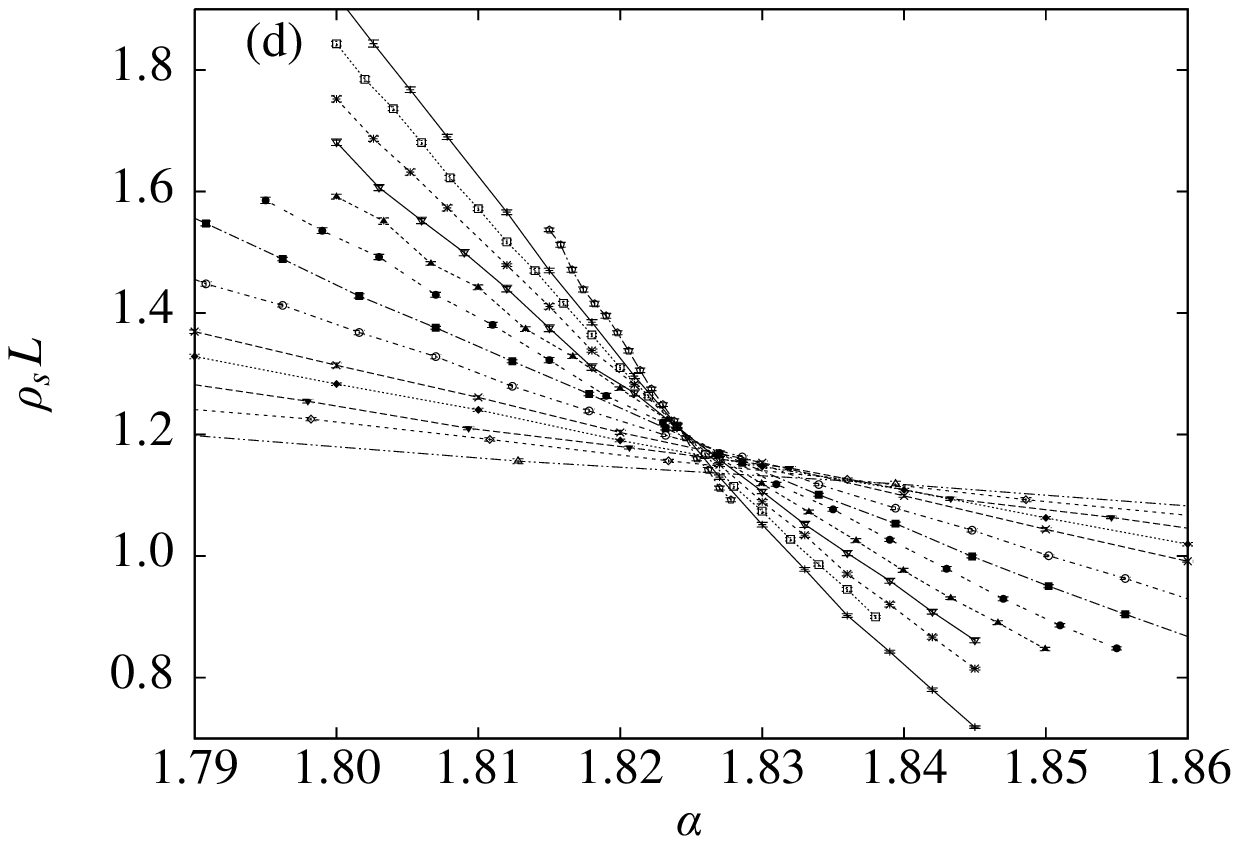}
\end{minipage}
\begin{minipage}{0.9\textwidth}
\includegraphics[width=\textwidth]{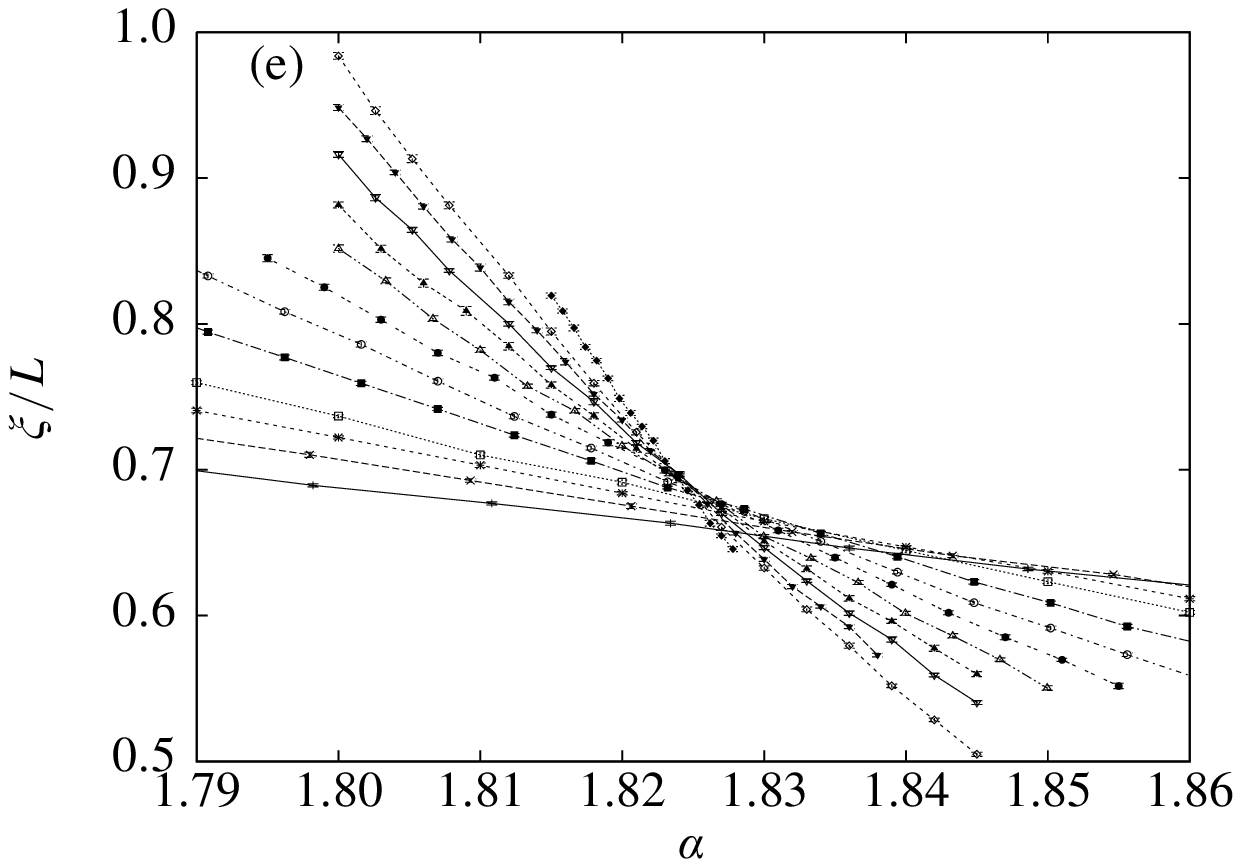}
\end{minipage}
\begin{minipage}{0.9\textwidth}
\includegraphics[width=\textwidth]{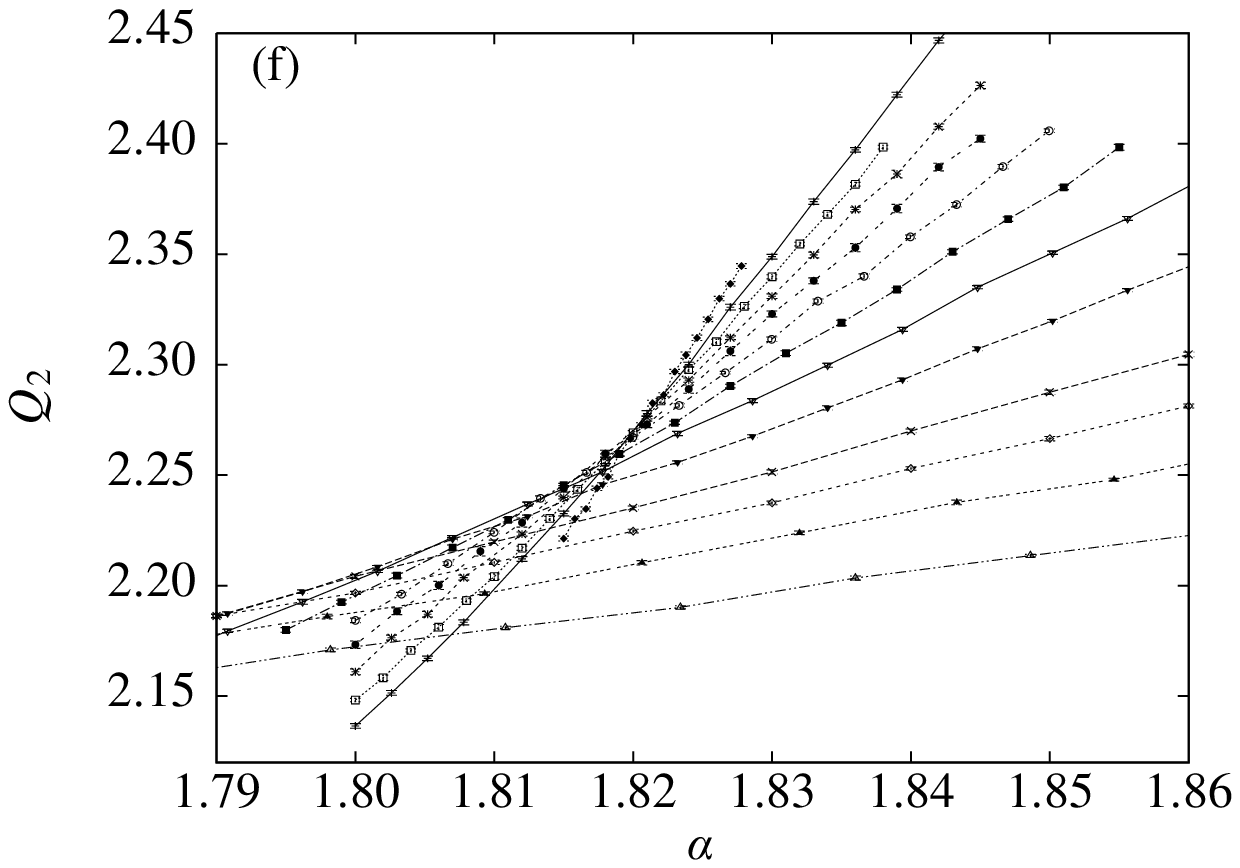}
\end{minipage}
\end{minipage}
\caption{\label{fig:observables} Behavior of different observables
  close to the quantum critical points. Different curves correspond to
  different system sizes (see Table~\ref{tab:parameters}), where
  larger slope means larger system size. The left column displays (a)
  the spin stiffness multiplied by the system size $L$, (b) the
  correlation length $\xi_y$ divided by system size $L$, and (c) the
  Binder parameter $Q_2$ for the ladder model. The plots (d)-(f) in
  the right column show the same quantities obtained for the plaquette
  model.}
\end{figure*}
\section{\label{sec:analysis} Simulation Results and The Critical Point}
We performed various simulations on the ladder and plaquette model
for lattice sizes specified in Table~\ref{tab:parameters} employing
the methods described in the last section. All runs where done using
periodic boundary conditions. The sample size of measured data is of
the order of $4\times 10^5$ for the plaquette model and $8\times 10^5$
in case of the ladder model, giving an indication that the ladder
model is somewhat harder to simulate. We typically performed $1\times
10^4$ sweeps for equilibration. Measurements were taken every sweep
and each sweep constructed as many loops as necessary in order to
visit $2n$ vertices in the SSE operator expansion on average.
\begin{table}[b]
\caption{\label{tab:parameters}Summary of lattice sizes $L$ studied in the simulations.}
\renewcommand{\arraystretch}{1.3}
\begin{tabularx}{\columnwidth}{Xl}
\hline\hline
 model  & lattice sizes $L$ \\
\hline
ladder & $8,10,12,14,16,20,24,28,32,36,40,52,64$ \\
plaquette & $8,10,12,16,20,24,28,32,36,40,44,48,56,72$ \\
\hline\hline
\end{tabularx}
\end{table}%
A summary of the raw data obtained from the simulations is displayed
in Fig.~\ref{fig:observables}, where we show the spin stiffness
$\rho_\mathrm{s}$, the correlation length $\xi$ and the Binder
parameter $Q_2$. The left and right panels in
Fig.~\ref{fig:observables} distinguish results for the ladder and
plaquette model, respectively.  Evidently, all quantities cross close
to an apparent quantum critical point justifying the scaling
assumptions for the observables described above.  However, clear
finite-size corrections can be observed for both cases as the crossing
points for small lattice sizes show large displacements. This behavior
is expected and in accordance to the data published in
Ref.~\onlinecite{wang:014431}. Our hope is that those corrections can
be described by the correction terms included in the scaling ansatz
\eqref{eqn:scaling4} (or \eqref{eqn:scaling5}). Using the raw data, we
will now try to extract a precise estimate of the quantum critical
point. To reach this aim, we will follow a two-stage process, starting
with an analysis of the crossing points followed by a finite-size
scaling investigation using the collapsing technique.

This will in principle also give us estimates of critical exponents
but we leave this issue for a more detailed investigation in
Sec.~\ref{sec:scaling} using ordinary and well-established methods.

\subsection{Estimation of the critical point from curve crossings}

\begin{figure}
\begin{minipage}{0.48\textwidth}
\includegraphics[width=0.9\textwidth]{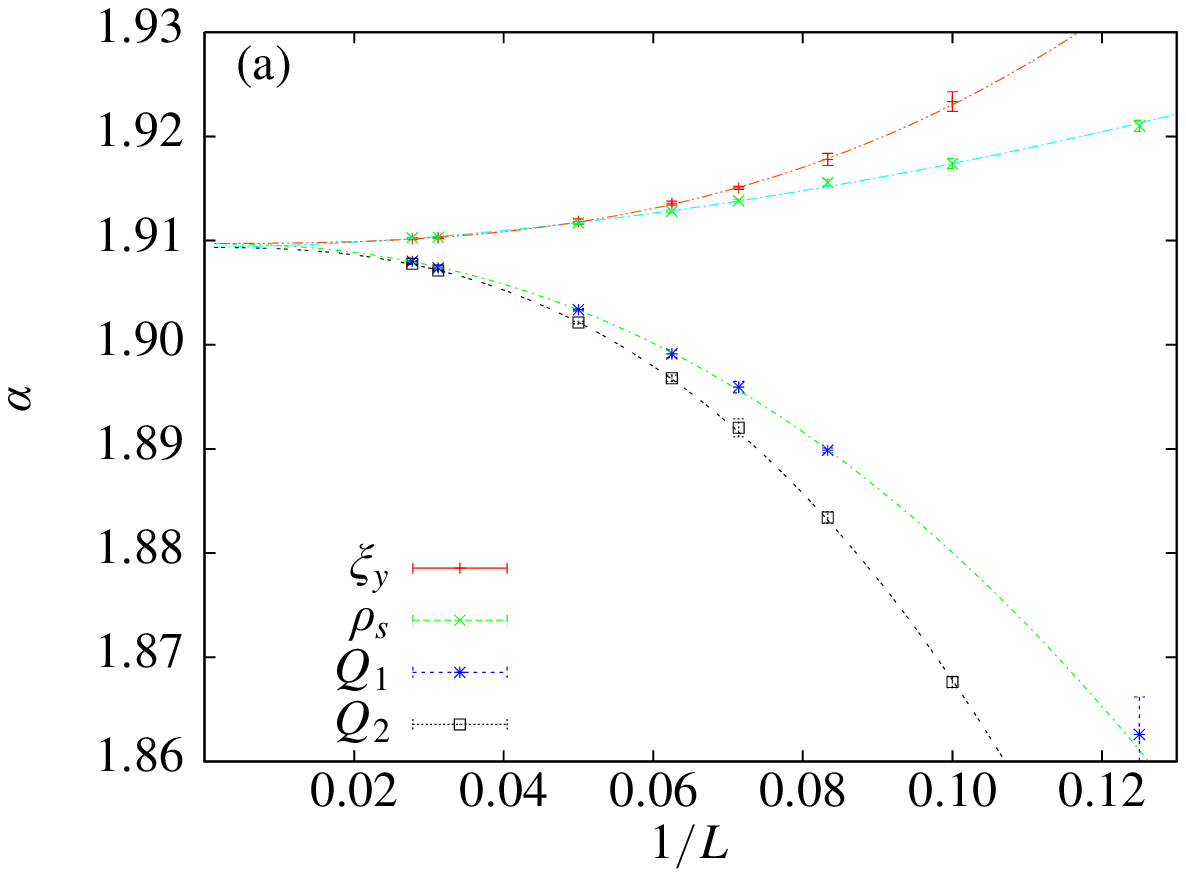}
\end{minipage}
\begin{minipage}{0.48\textwidth}
\includegraphics[width=0.9\textwidth]{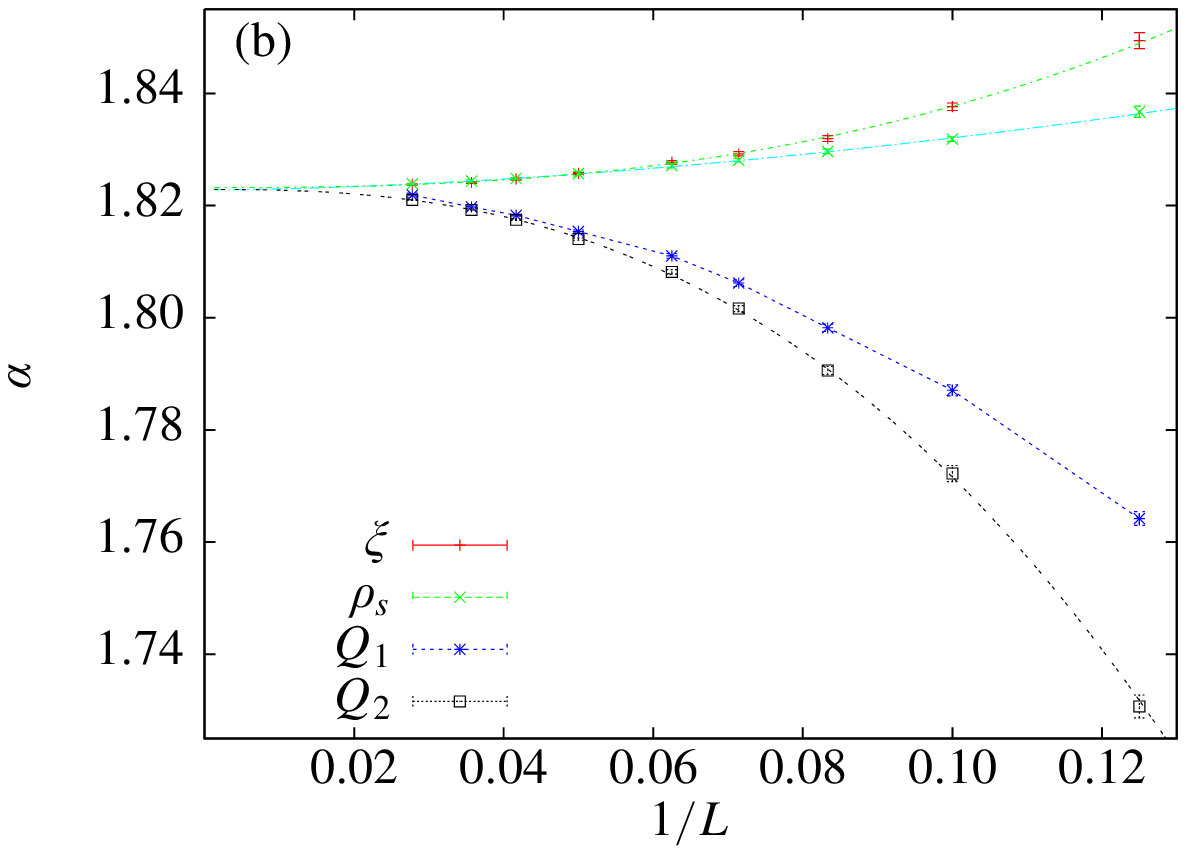}
\end{minipage}
\caption{\label{fig:crossing_scaling}(Color online) Crossing points
  from data at $L$ and $2L$ of the four quantities $\rho_sL$,
  $\xi/L$, $Q_1$, and $Q_2$ versus the inverse lattice size. At
  $L=\infty$ all curves should meet and define the quantum critical
  point $\alpha_c$. (a) Analysis for the ladder model.
  (b) Same for the plaquette model.}
\end{figure}

Finite-size scaling analysis with scaling functions involving many
free parameters is a tedious and difficult task due to well-known problems of multidimensional nonlinear minimization. Before we attempt to
perform a full finite-size scaling study using
Eq.~\eqref{eqn:scaling4}, we would therefore like to set bounds on the
possible values of the critical coupling $\alpha_c$. To this end, a
convenient approach consists in looking at the scaling of crossing
points of curves at $L_1$ and $L_2$ (where $L_2=2\,L_1$) for different
values of $L_1$. The crossing points are easily obtained using either
the multihistogram method or fitting data at $L_1$ and $L_2$ to the
simple scaling ansatz in Eq.~\eqref{eqn:scaling1} (using polynomial
interpolation). Performing this procedure on the various observables
of Fig.~\ref{fig:observables} (and $Q_1$) yields the plots of
Fig.~\ref{fig:crossing_scaling}, which show convergence of the
intersection points to the quantum critical point in the thermodynamic
limit.  The plots are presented with an $x$-axis as $1/L$, since we do
not know the correct scaling \textit{a-priori}.
The qualitative behavior of the different quantities toward the
critical point is rather similar to Ref.~\onlinecite{wang:014431}. We
find, that for both models the spin stiffness has the least
finite-size corrections, followed by the correlation length, and that
the normal Binder parameter $Q_2$ shows large deviations at small
lattice sizes. This is not necessarily a disadvantage since this often
leads to better controlled fits. Before performing some fits, however,
let us emphasize that in case of the staggered model considered in
Ref.~\onlinecite{wenzel-JJp-2008}, the spin stiffness displayed a
qualitatively different convergence toward the infinite-volume limit
because there the correlation length $\xi_y$ showed less finite-size
effects.  This proves that $\rho_\mathrm{s}$ is not always the best
quantity.

Since all quantities give a rather consistent picture in their scaling
properties we can safely bracket the critical couplings from the
crossings using the largest available $(L,2L)$ pair. This yields
$\alpha_\mathrm{c}\in [1.9070,1.9105]$ and $\alpha_\mathrm{c}\in
[1.821,1.834]$ for the ladder and plaquette cases, respectively. It is
tempting to obtain a more precise estimate from fitting the crossing
points to an ansatz due to Binder\cite{Binder1981}
\begin{equation}\label{eqn:fitcross}
  \alpha_\mathrm{c}(L,2L)=\alpha_\mathrm{c}+\frac{b}{L^{1/\nu+\omega}}\,,\end{equation} which
states that the crossings should normally converge faster than
$L^{-1/\nu}$, and would indeed show no $L$-dependence at all if
$\omega=\infty$, i.e. no correction. In this ansatz $b$ is a constant and we neglected
subleading corrections from the ``shift'' term $\phi$. This term can in principle be included,\cite{beach-2005} leading to fits
which are more difficult to perform. The smooth curves in
Fig.~\ref{fig:crossing_scaling}(a) for the ladder model correspond to fits for the
correlation length, the spin stiffness, and the Binder parameters
$Q_1$ and $Q_2$, which yield $\alpha_\mathrm{c} = 1.9097(3)$ ($\xi_y$), $\alpha_\mathrm{c} = 1.9092(6)$
($\rho_\mathrm{s}$), $\alpha_\mathrm{c} =1.9095(5)$ ($Q_1$), and  $\alpha_\mathrm{c} =1.9093(3)$ ($Q_2$). They are all in agreement within error bars. For the plaquette model
(Fig.~\ref{fig:crossing_scaling}(b)) we obtain in the same order $\alpha_\mathrm{c}=1.8232(3)$, $\alpha_\mathrm{c}=1.8228(4)$, $\alpha_\mathrm{c} =1.8238(8)$, and
$\alpha_\mathrm{c}=1.8229(4)$, respectively. All fit results are
summarized in Table~\ref{tab:fitcrossings}, where we additionally give the fitted quantity $1/
\nu+\omega$ and the quality of the fits through the chi-squared per degree of freedom ($\chi^2/\mathrm{d.o.f.}$). Under the assumption that the correlation length exponent $\nu \approx 0.7$, we deduce
that $\omega$ lies roughly in the interval $[0.8,1.2]$ for the correlation length and the Binder parameters. For the spin stiffness, interestingly, 
$\omega$ seems to be smaller. The stiffness thus appears to cross close to the quantum critical point but has slow convergence towards it. On the other hand, the spin stiffness could not be well described by Eq.~\eqref{eqn:fitcross}. 
A similar effect will, in fact, be seen in the analysis of Sec.~\ref{sec:collapse}. 

We feel that the critical points obtained above give a fair estimate as
they agree within error bars. A posteriori, this justifies the
neglection of $\phi$.  Finally, it should be clear, that by the same
approach other estimates, like $\nu$ and $g_0$, can and have been
bracketed aiding in the collapse analysis now to come.
\begin{table}[t]
\caption{\label{tab:fitcrossings} Estimates for the critical point derived from Eq.~\eqref{eqn:fitcross} for the ladder (top group) and plaquette model (bottom group).}
\renewcommand{\arraystretch}{1.4}
\begin{tabularx}{\columnwidth}{XXXX}
\hline\hline
quantity & $\alpha_\mathrm{c}$ & $1/
\nu+\omega$ & $\chi^2/\mathrm{d.o.f}$ \\
\hline
$\xi_y$ & $1.9097(3)$ & $2.6(1)$ & $0.85$ \\
$\rho_\mathrm{s}$ & $1.9092(6)$ & $1.7(2)$  & $1.7$ \\
$Q_1$ & $1.9095(5)$ & $2.3(1)$ & $1.2$\\
$Q_2$ & $1.9093(3)$ & $2.55(8)$ & $0.64$ \\\hline
$\xi$  & $1.8232(3)$ & $2.6(1)$ & $0.24$ \\
$\rho_\mathrm{s}$ & $1.8228(4)$ & $1.8(1)$ & $0.14$ \\
$Q_1$ & $1.8238(8)$ & $2.2(1)$ & $0.82$ \\
$Q_2$ & $1.8229(4)$ & $2.6(1)$ & $0.72$ \\
\hline\hline
\end{tabularx}
\end{table}%

\subsection{\label{sec:collapse} The critical point from data collapses}
In the previous section first estimates of the critical points were
obtained. Next, our goal is to cross-check and possibly improve the
accuracy by analyzing the data for the full set of $\alpha$ values
around the crossing points including all lattice sizes in
Table~\ref{tab:parameters}.  We will therefore now elaborate on the
data collapse procedure to the scaling ansatz of
Eq.~\eqref{eqn:scaling4}, knowing that we have to include subleading
corrections terms. In this process we will leave all parameters free,
since we want to avoid preoccupation about the universality class. Of
course we keep in mind the bracketing of some important quantities in
the previous section.  Fitting is done using Eq.~\eqref{eqn:expansion}
or \eqref{eqn:collapsequality}. The two approaches have been compared
and we could not detect a noticeable difference in the outcome. We
hence use the less time consuming approach according to
Eq.~\eqref{eqn:expansion} for which a fourth-order polynomial for
$g_\mathcal{O}(x)$ is employed.

Due to potential problems with multidimensional fitting, the analysis is repeated
for at least two different scenarios. In a first case we
ignore the subleading shift correction, i.e.\ we set
$\phi=\infty$ (or $d=0$) to obtain a first idea of the critical coupling, the correlation
length exponent $\nu$ and other parameters. We will see that apart
from a few exceptions, this approach actually describes our data well
enough.  Next we repeat the collapse taking into account possible
shift corrections, described by a finite $\phi$.
\begin{figure}[b]
\begin{minipage}{0.8\columnwidth}
\includegraphics[width=\textwidth]{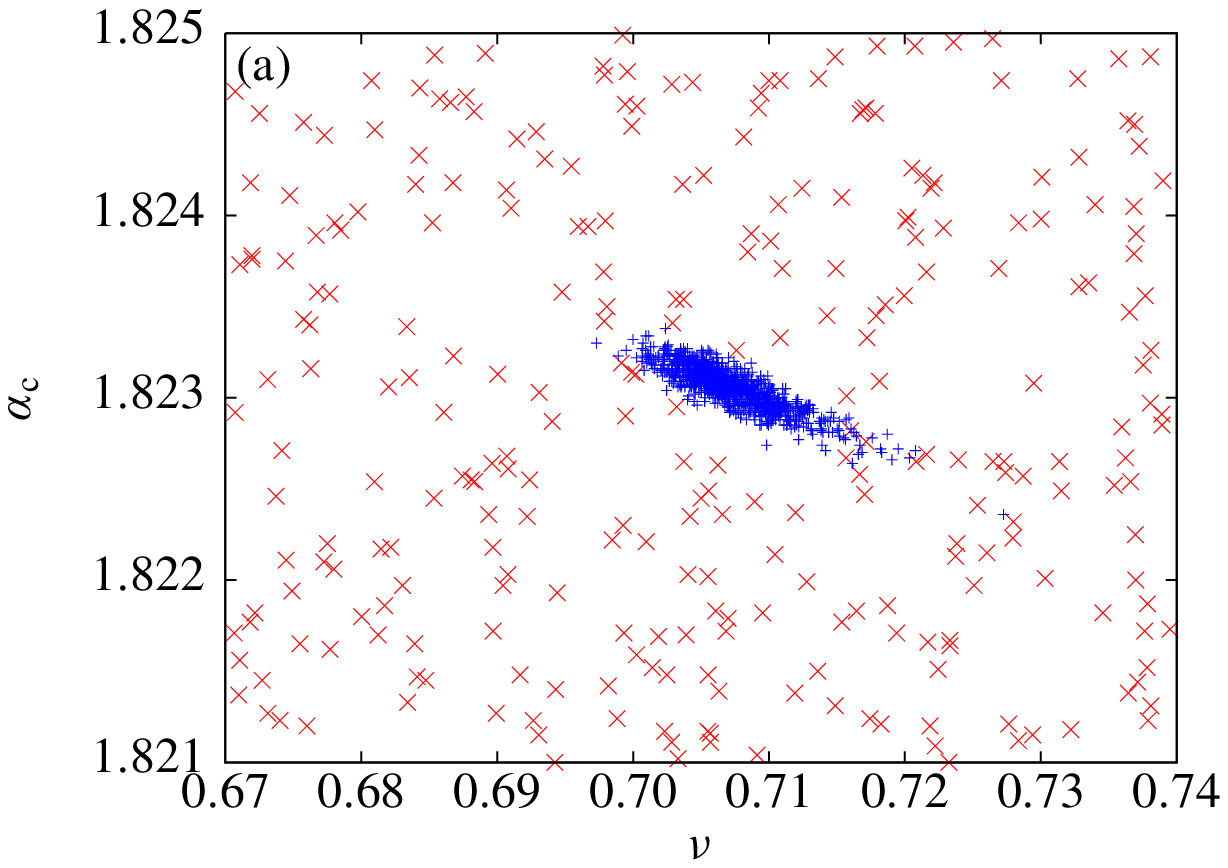}
\end{minipage}
\begin{minipage}{0.49\columnwidth}
\includegraphics[width=\textwidth]{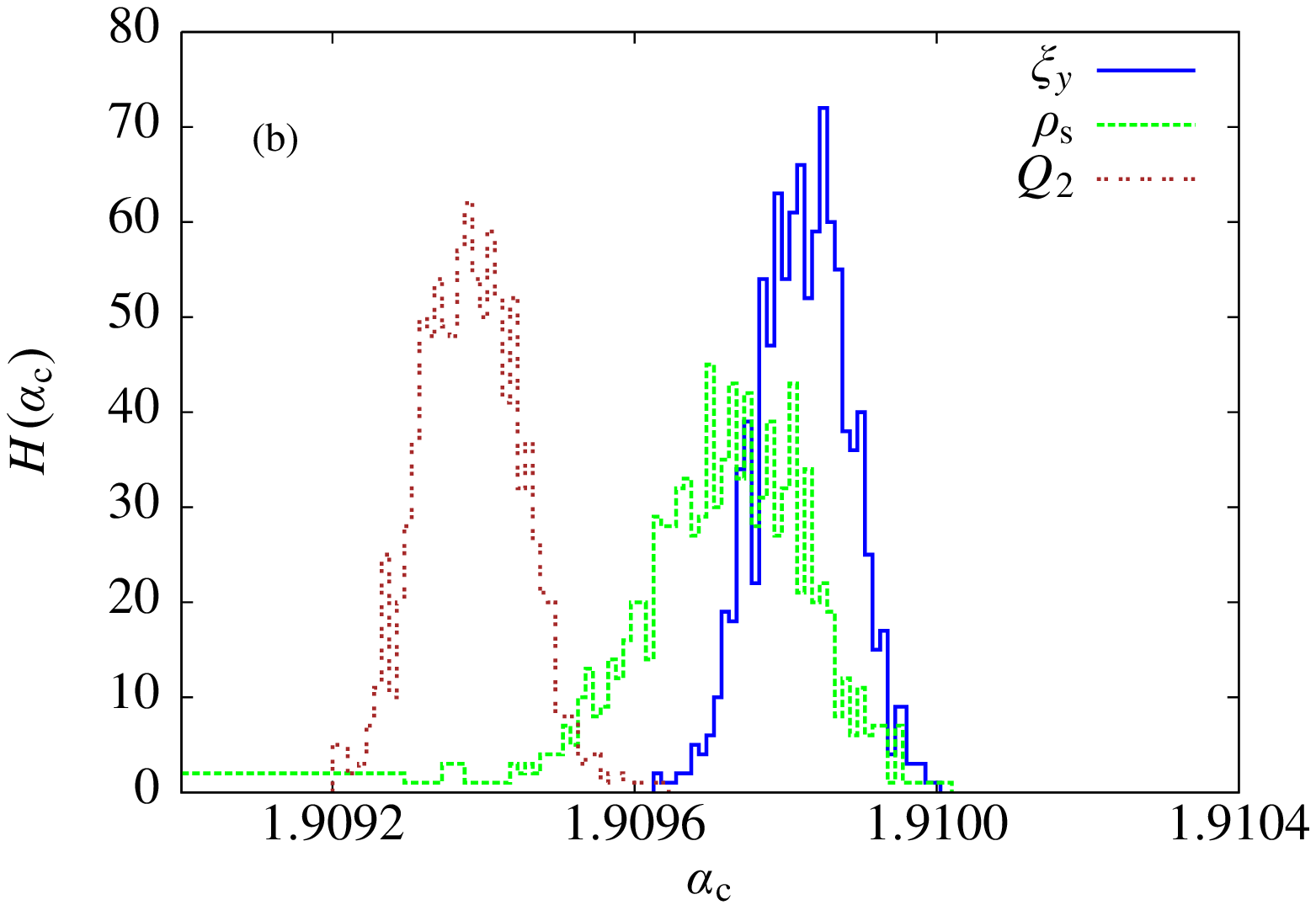}
\end{minipage}
\begin{minipage}{0.49\columnwidth}
\includegraphics[width=\textwidth]{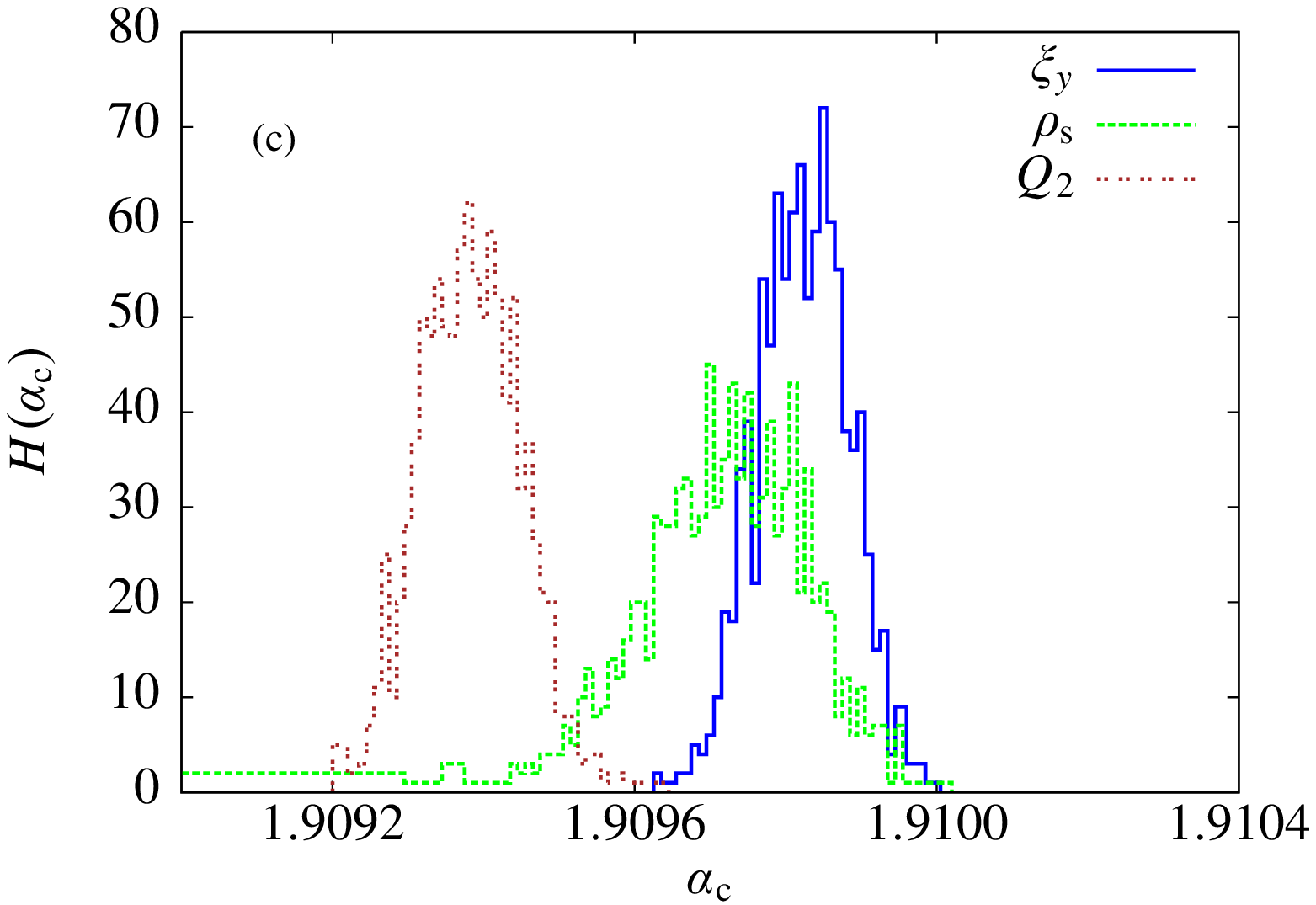}
\end{minipage}
\caption{\label{fig:gc_histos}(Color online) (a) Visualization of the
  collapsing analysis. The (red) crosses ($\times$) signify the starting values and
  dense (blue) points ($+$) the final values of ($\alpha_\mathrm{c}$, $\nu$) of the
  procedure for the spin stiffness in case of the plaquette model.
  Histograms of the final value for $\alpha_\mathrm{c}$ for several
  observables for (b) the ladder model, and (c) the plaquette model.}
\end{figure}%%
%%
%% this table summarizes results from the collapsing procedure 
%% for the plaquette and ladder model
%%
\begin{table}
\renewcommand{\arraystretch}{1.4}
\caption{\label{tab:result} Tabulated results for the critical coupling ratio $\alpha_\mathrm{c}$, the exponent $\nu$, and the factor $g_0$ from the collapse procedure for both the ladder (upper group) and the plaquette model (lower group). In some cases results from two fits, with and without a $\phi$ term are given.}
\begin{tabularx}{\columnwidth}{X l l l l}
\hline\hline 
& restr. & $\alpha_c$ & $\nu$ & $g_0$\\
\hline
$Q_2$ & $d=0$ & $1.9094(3)$  & $0.717(10)$ &  $2.32(1)$\\
$Q_1$ & $d=0$ & $1.9096(4)$  & $0.72(1)$  &   $1.451(8)$\\
      & no &    $1.9094(3)$  & $0.72(1)$  &   $1.449(3)$\\
$\rho_\mathrm{s}L$ &  no & $1.90974(15)$ & $0.705(7)$ & $1.155(10)$\\
$\xi_y/L$ & $d=0$ & $1.9098(4)$ & $0.715(10)$  &  $0.62(1)$\\
\hline
$Q_2$ & $d=0$ & $1.8228(4)$  & $0.716(6)$  &  $2.313(6)$\\
       & no    & $1.8227(4)$  & $0.72(1)$ &  $2.311(5)$\\
$Q_1$ & $d=0$ & $1.8238(6)$  & $0.72(1)$  &  $1.453(2)$\\
       & no    & $1.8228(6)$  & $0.72(1)$ &  $1.447(4)$\\
$\rho_\mathrm{s}L$ & $d=0$ & $1.8230(3)$  & $0.67(1)$  &  $1.28(3)$\\
       & no    & $1.8230(2)$  & $0.707(6)$ &  $1.27(2)$\\
$\xi/L$  & $d=0$ & $1.8232(2)$ & $0.709(6)$  &  $0.706(5)$\\
          & no   & $1.8231(2)$ & $0.713(6)$  &  $0.70(1)$\\
\hline\hline
\end{tabularx}%
\end{table}%
All fits are repeated multiple times including random noise on the
starting parameters as well as on the raw data. In the latter case,
the noise is taken to be normal distributed and within the
Jackknife\cite{efron} errors $\sigma$ of the original data points. We
typically perform $1000$ fits for each observable. All quoted error
bars are then understood as being the error bars from this
bootstrap\cite{efron} procedure. Figure~\ref{fig:gc_histos}(a)
outlines this procedure and shows that the collapse is well behaved.
Random starting values converge to a narrow collapse region.
Figures~\ref{fig:gc_histos}(b,c) display histograms of the final
critical couplings obtained from the bootstrap procedure for the
ladder and plaquette model, respectively.  It is seen that the results
are consistent as they more or less overlap, yet we note a systematic
effect as the Binder parameter tends to give smaller estimates in
comparison to the correlation length and the spin stiffness.  This is
also in accordance with the data on the full bilayer of
Ref.~\onlinecite{wang:014431}.  Table~\ref{tab:result} summarizes
concrete results for the different models and observables.  The best
results for $\alpha_\mathrm{c}$ are obtained from the spin stiffness
which usually interpolates between values from the correlation length
and $Q_2$.

Second, we could not detect a noticeable difference in the results if
we include a $\phi$ degree of freedom. An exception to this
observation is the spin stiffness, which showed controlled fits only
in presence of $\phi$ (which probably acts as a kind of stabilizer).
This fact agrees with the observation made during the analysis of the
crossing points above but is presently not well understood. The results for the exponent $\nu$ are
consistent with $O(3)$ universality.  Finally, typical values for
$\omega$ are in the range of $\omega\in [0.8,1.3]$, consistent with
the previous section. In case of the spin stiffness, we obtain
$\omega\approx 1.4 $ and $\phi/\nu \approx 2.5$.
\begin{figure}[t]
\centering
\includegraphics[width=0.8\columnwidth]{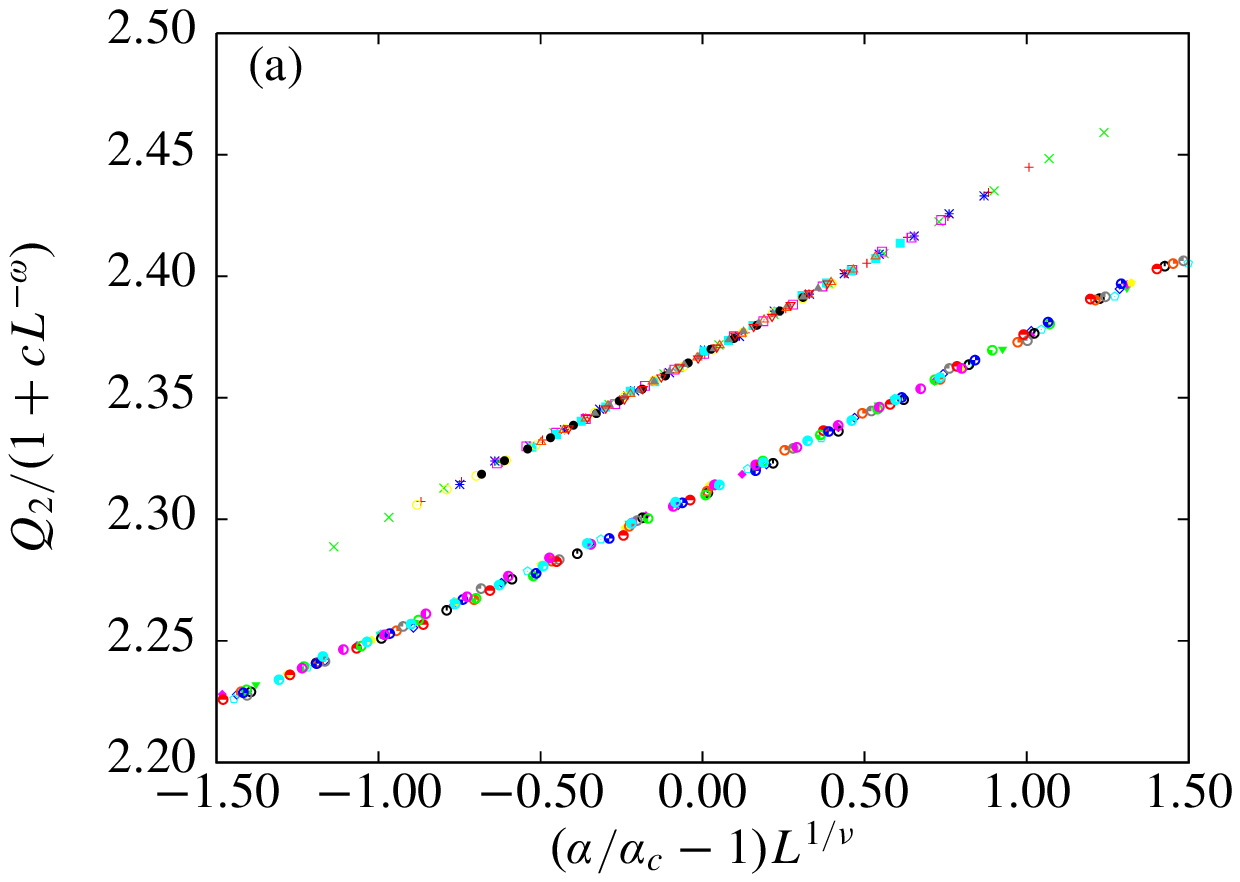}
\includegraphics[width=0.8\columnwidth]{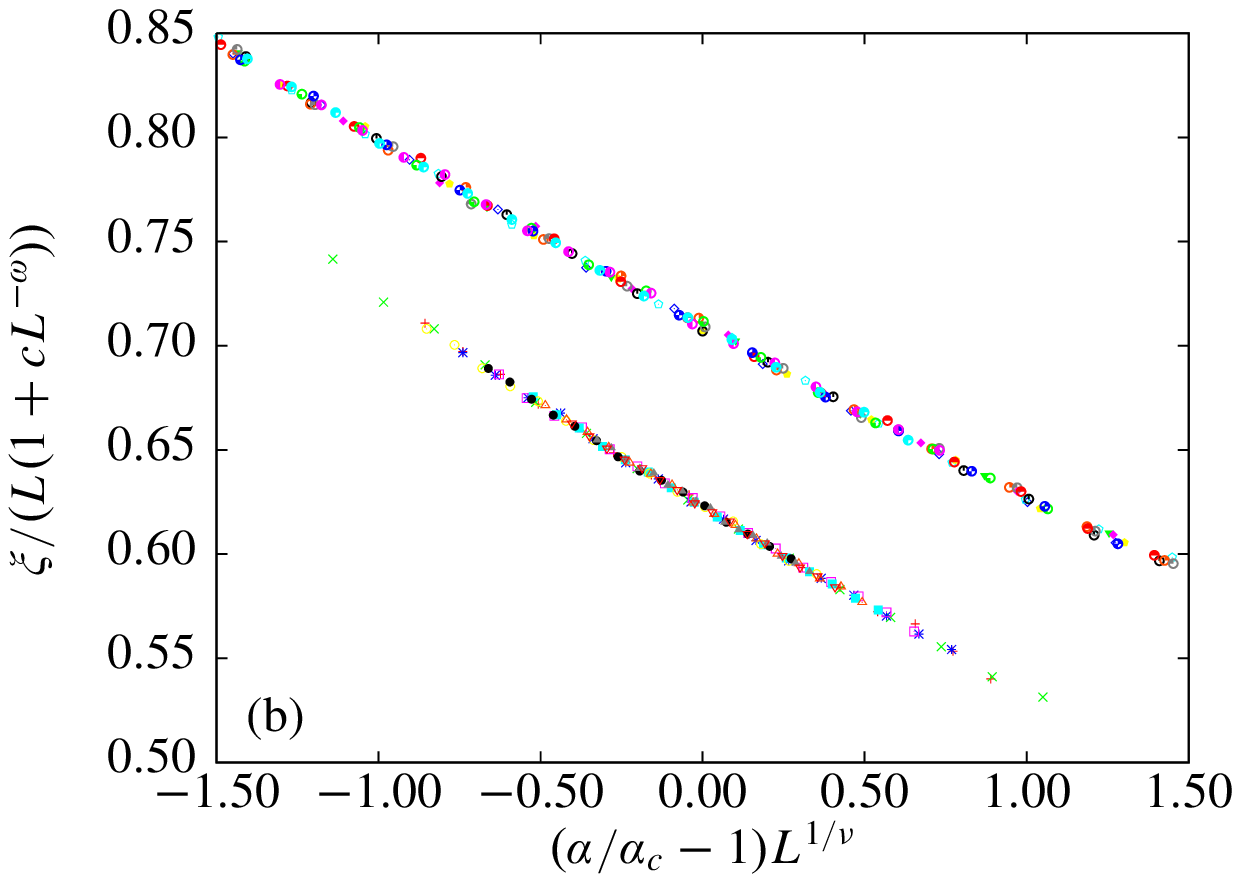}
\includegraphics[width=0.8\columnwidth]{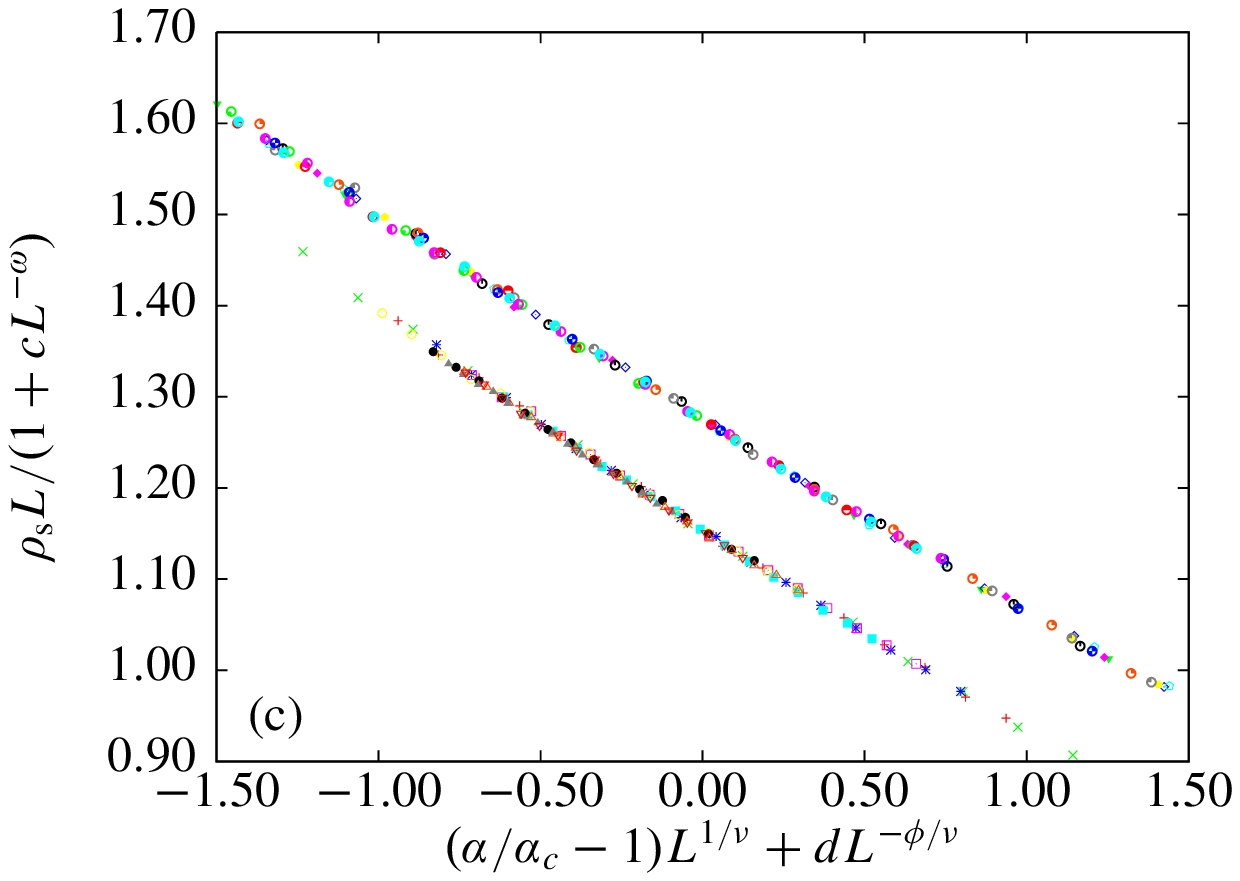}
\caption{\label{fig:collapses}(Color online) Data collapses for the
  ladder and plaquette (wider range) model displaying (a) the Binder
  ratio $Q_2$ (where the collapse for the ladder model was shifted
  upwards by $0.05$ for better visibility), (b) the correlation length
  $\xi_y$ respectively $\xi$, and (c) the spin stiffness
  $\rho_\mathrm{s}$.  Apart from the special case of the spin stiffness, collapses are shown without the the $\phi$ correction term.}
\end{figure}%
Using these results, concrete data collapses of the original data are
given in Fig.~\ref{fig:collapses} which display a very good collapse
quality.

In principle, one would need to perform additional investigations on
the influence of size of the collapsed regime $x$ (see
Eq.~\eqref{eqn:collapsequality}). We have done that partly, but do not
attempt a detailed extrapolation as we will extract the actual
critical exponents by a different method. In any case, we believe that
our estimates for $\alpha_\mathrm{c}$ are correct beyond doubt as they
are consistently obtained from three independent methods (crossing
analysis, collapse to \eqref{eqn:scaling4}, and collapse to
\eqref{eqn:scaling5}). This also justifies the use of the
approximations which are present in the finite-size scaling ansatz.

We now state the main result of this section in giving our final
estimates for the critical couplings. Since no details about
systematic errors (e.g. from undescribed correction effects etc.) are
known, a plain average of the critical coupling estimates from $Q_2$,
$\rho_\mathrm{s}$, and $\xi_y$ is probably the best choice (and is the
same as a weighted average). This way, our final estimate is
$\alpha_\mathrm{c}=1.9096(2)$ and $\alpha_\mathrm{c}=1.8230(2)$ for
the ladder and plaquette models, respectively. In case of the ladder
model, this result is in slight disagreement with the previous value
of $1.9107(2)$ in Ref.~\onlinecite{PhysRevB.65.014407}.

Before we go on, it is interesting to observe from the quantities
$g_0$ listed in Table~\ref{tab:result}, that both the Binder
parameters at the crossing point seem to be consistent within error
bars among the two models, whereas the spin stiffness and the
correlation length clearly do not possess this property but the reader
should keep in mind that $\xi_y$ and $\xi$ are slightly different
quantities. %This
%further demonstrates the pseudo-universality of the Binder parameters
%given the same boundary conditions.

\section{\label{sec:scaling}Scaling at criticality}

Having determined estimates for the critical couplings, we now turn to
an investigation of the critical exponents. To this end, we make use of
standard methods of Monte Carlo data analysis. Our reason to decouple this
investigation from the collapse analysis is to get independent and
unbiased estimates. A fit at a predetermined critical point, secondly,
has less degrees of freedom and is hence easier to control.

Analysis of the exponents is performed using standard relations
and definitions. An established method to obtain the correlation length
exponent $\nu$, is via the slope
$s_{Q_2}=\mathrm{d}Q_2/\mathrm{d\alpha}$ of the Binder parameter
evaluated at the critical point. Using Eq.~\eqref{eqn:scaling2} we
arrive at
\begin{equation}
\label{eqn:slopeBinder}
s_{Q_2}\sim L^{1/\nu}\,.
\end{equation}
Other exponents, in particular, $\beta$ and $\eta$ are calculated from the order parameter and the structure factor at criticality as
\begin{equation}
  \label{eqn:beta}
  \langle |m^z_s| \rangle\sim L^{-\beta/\nu}\,,\quad S(\pi,\pi)\sim L^{1-\eta}\,,
\end{equation}
where we assume Lorentz invariance, i.e., $z=1$ in $S(\pi,\pi)\sim
L^{2-z-\eta}$.

In order to use Eqs.~\eqref{eqn:slopeBinder}, \eqref{eqn:beta} we need
data at the quantum critical point.  This can in principle be achieved
by performing new simulations. Since we have rather good data in the
vicinity of $\alpha_\mathrm{c}$ already, we instead choose to compute
$s_{Q_2}$, $m^z_s$, and $S(\pi,\pi)$ from polynomial interpolation or
multihistogram reweighting as employed in the last section. We have
checked the consistency of the two approaches and use the first method
from here on.  Again, a bootstrap with $1000$ samples is performed on
top of this interpolation, varying the raw input data within the
uncertainties.
\begin{figure}[t]
\centering
\begin{minipage}{\columnwidth}
\includegraphics[width=0.8\textwidth]{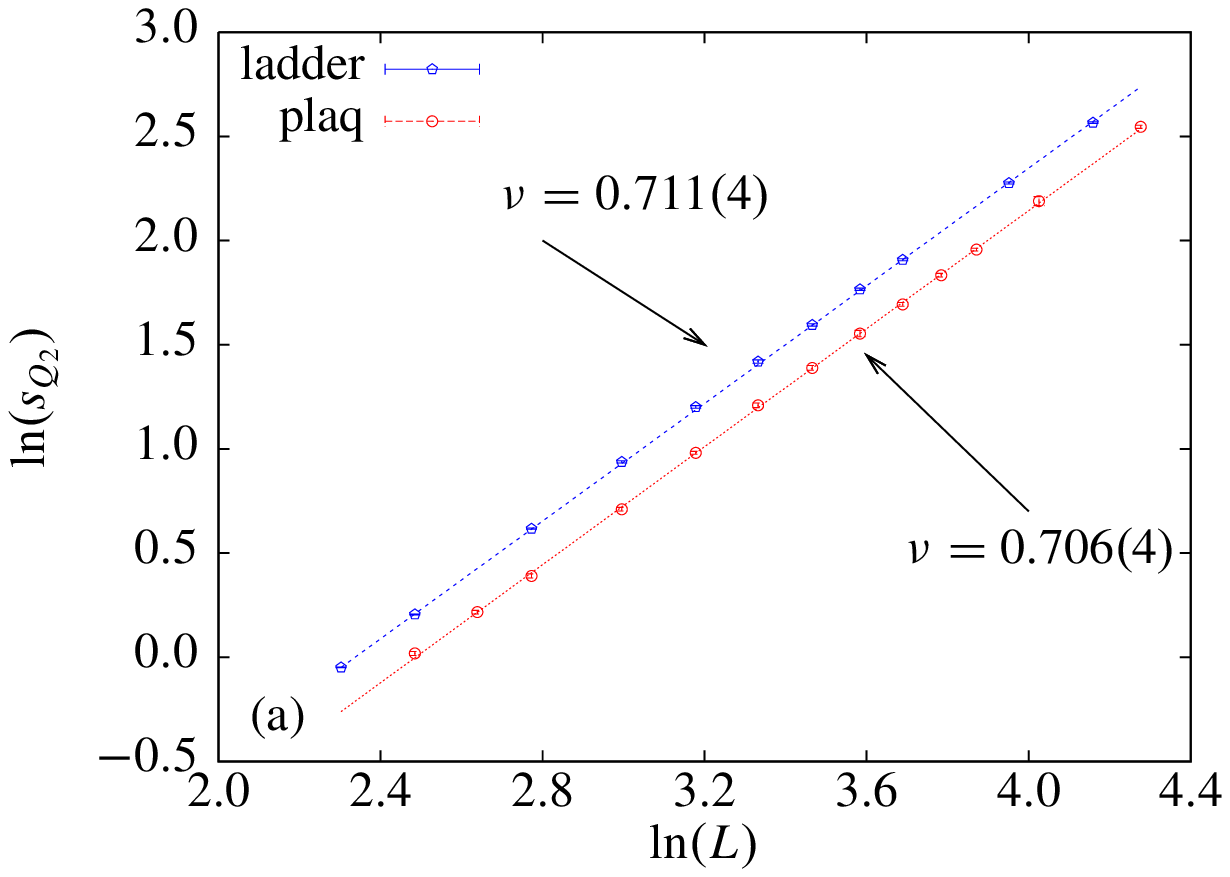}
\end{minipage}
\begin{minipage}{\columnwidth}
\includegraphics[width=0.8\textwidth]{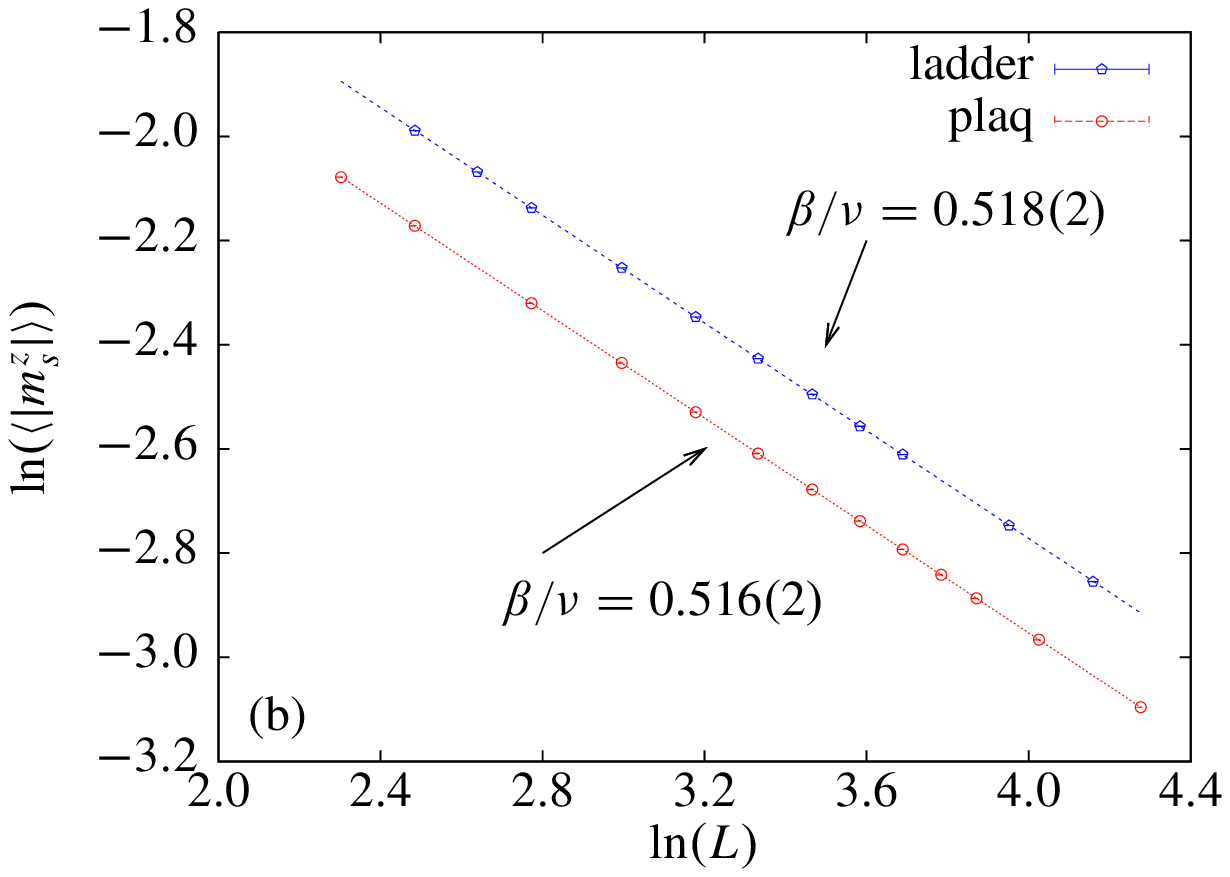}
\end{minipage}
\begin{minipage}{\columnwidth}
\includegraphics[width=0.8\textwidth]{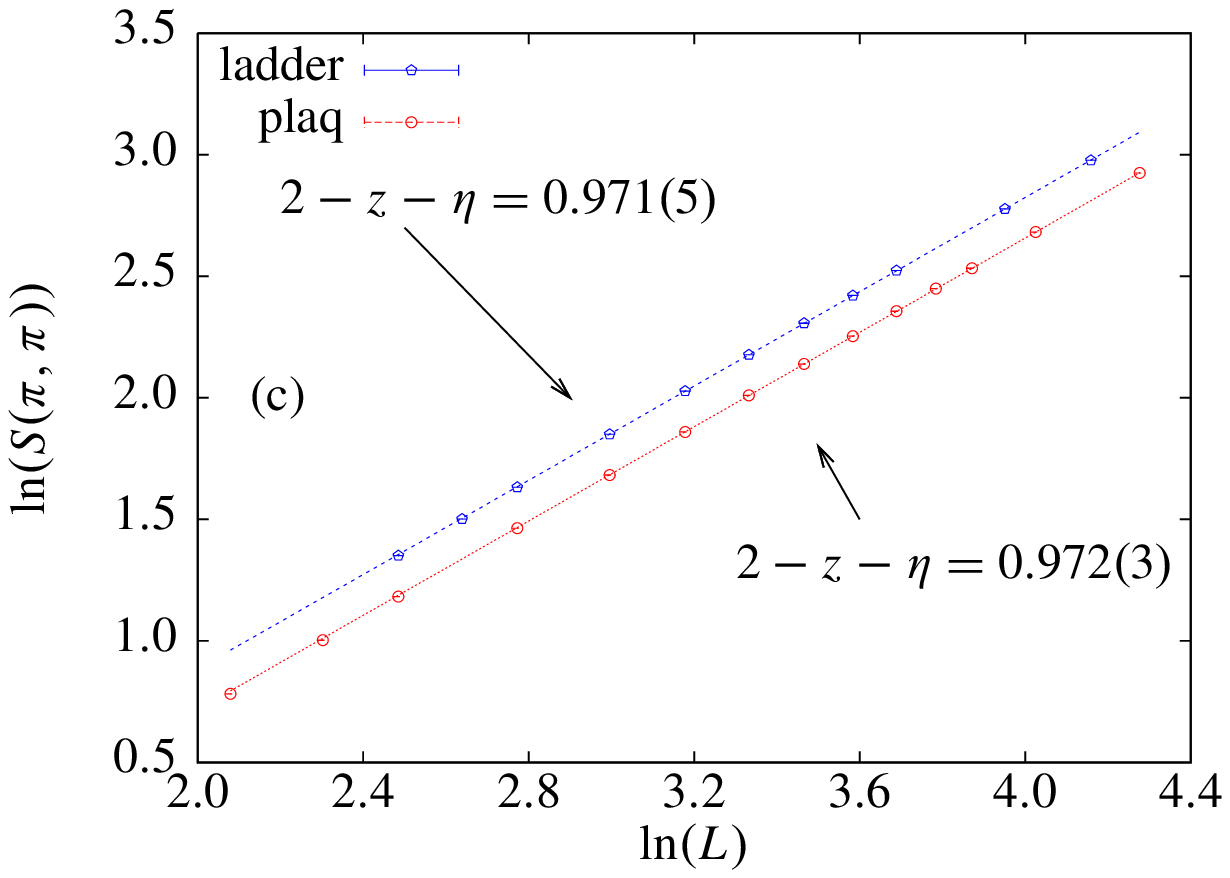}
\end{minipage}
\caption{\label{fig:smcmp} (Color online) Finite-size scaling using
  (a) the slope $s_{Q_2}$ of the Binder parameter, (b) the staggered magnetization, and (c) the staggered
  structure factor $S(\pi,\pi)$. These quantities are computed at the
  critical points determined in Sec.~\ref{sec:analysis}.}
\end{figure}%
Figure~\ref{fig:smcmp} summarizes and displays the critical data so
obtained. All plots are in a $\ln-\ln$ style vs. the lattice size $L$. It is evident
that straight lines represent the data rather well. To make this
statement more quantitative we now perform and present detailed fits
and their results in Table~\ref{tab:fitting}.
\begin{table}
  \caption{\label{tab:fitting}Fit results for the critical exponents $\nu$, $\beta/\nu$, and $\eta$. We summarize results including a variation of the critical point within its error bar. For the ladder model (top group of values) fit results and quality of fits are also given at the previous best estimate of $\alpha_\mathrm{c}$. The bottom group are results for the plaquette model. Numbers in $[\ldots]$ brackets denote the $\chi^2/\mathrm{d.o.f}$. For comparison relevant reference values for the 3D $O(3)$ universality class are given in the last line.}
  \renewcommand{\arraystretch}{1.4}
  \begin{tabularx}{\columnwidth}{lXXX}
    \hline\hline
    $\alpha_\mathrm{c}$ & $\nu$\footnote{$L>12$.} &  $\beta/\nu$\footnote{$L>16$.} & $\eta$\footnote{$L>20$.} \\
    \hline
    $1.9096 - \sigma$ & $0.712(4)$ $[1.8]$ &  $0.516(2)$ $[0.5]$ & $0.026(2)$ $[0.2]$ \\
    $1.9096$          & $0.711(4)$ $[1.8]$ &  $0.518(2)$ $[1.1]$ & $0.029(5)$ $[0.8]$ \\
    $1.9096 + \sigma$ & $0.710(4)$ $[1.8]$ &  $0.519(3)$ $[2.5]$ & $0.032(7)$ $[1.4]$ \\
    $1.9107$\footnote{Previous best estimate of Ref.~\onlinecite{PhysRevB.65.014407}.} & $0.709(3)$ $[1.7]$ & $0.525(8)$ $[15.3]$ & $0.051(10)$ $[12]$ \\\hline
    $1.8230 - \sigma$ & $0.708(4)$ $[0.99]$  & $0.515(2)$ $[0.84]$ & $0.025(4)$ $[0.15]$ \\
    $1.8230$ & $0.706(4)$ $[1.04]$  & $0.516(2)$ $[0.40]$ & $0.028(3)$ $[0.31]$ \\
    $1.8230 + \sigma$ & $0.706(4)$ $[1.10]$  & $0.517(2)$ $[1.6]$  & $0.031(5)$ $[0.80]$ \\
    Ref.~\onlinecite{hasenbuschO3} & $0.7112(5)$ & $0.5188(3)$  & $0.0375(5)$ \\
    \hline\hline
  \end{tabularx}
\end{table}%
% (note for plaquette model took 3rd order fits for $Q_2$).
For each quantity, 3 fits are performed corresponding to the best
estimate of $\alpha_\mathrm{c}$, as well as its lower and upper bounds
from the uncertainty. In case of the ladder model, we also try a
further fit at the previous estimate of
Ref.~\onlinecite{PhysRevB.65.014407}. Several observations can be made
regarding our results. First, the exponent $\nu$ obtained from the
slope of the Binder parameter is rather insensitive to the variation
in $\alpha_\mathrm{c}$. Medium to good quality fits can be performed
for lattice sizes $L>12$ for both models. All results for $\nu$ are
consistent with the best known value $0.7112(5)$ for the 3D $O(3)$ universality class.\cite{hasenbuschO3} Our estimate for
$\nu$ as in Table \ref{tab:fitting} improves the accuracy compared to
Ref.~\onlinecite{PhysRevB.65.014407} by about one order of magnitude.
However, we do not quite reach the level presented for the bilayer
models.\cite{wang:014431} This could be related to the more
complicated nature of the phase transition in planar models, where
in-plane symmetries are broken.

In case of the exponent $\beta/\nu$, good fits to
Eq.~\eqref{eqn:beta} could be performed for $L>16$ resulting in almost perfect agreement
with the reference value of $\beta/\nu=0.5188(3)$, which we computed
from Ref.~\onlinecite{hasenbuschO3}. Note that in case of the ladder
model, however, the $\chi^2/\mathrm{d.o.f}$ increases by one order
of magnitude accompanied by an increase of the value of $\beta/\nu$ when performing
the fit at the previous estimate for $\alpha_\mathrm{c}$. This
indicates that the result of this paper indeed captures the critical point
in the ladder model more accurately.  The same observation is true for the exponent $\eta$.
All results for this exponent are quoted for
lattice sizes $L>20$, indicating that this quantity is harder to
estimate. Yet, our results are still consistent or close to the
reference value. A natural check on the consistency of our results is
a test of the (hyper)scaling relation $2\beta/\nu=(d+z-2+\eta)$, which
seems to be satisfied for nearly all cases, but it is also clear that
$\eta$ and $\beta$ are probably strongly correlated as they derive
from almost the same quantity.

Finally, the interested reader is referred to
Ref.~\onlinecite{wenzel-JJp-2008} for a slight extension of the
current scaling analysis. In that reference a further comparison
regarding the Binder parameter at the critical point in different
planar and bilayer Heisenberg models is presented.

\section{\label{sec:conclusions} Summary and Conclusions}
In conclusion, we have considered two particular geometric
arrangements of competing interactions in 2D planar quantum Heisenberg
models complementing work we have started in
Ref.~\onlinecite{wenzel-JJp-2008}. From detailed QMC simulations and a
finite-size scaling study, this work provides a first high-precision
value for the critical point in the plaquettized Heisenberg model and
improves the value for the ladder model. In both cases, the use of
correction terms and a combined analysis of different quantities is
essential. For both models we derive the full set of critical
exponents and improve their accuracy by about one order of magnitude
(from $\nu=0.71(3)$ to $0.711(4)$) for the ladder model. These values
are in excellent agreement with the classical 3D $O(3)$ universality
class.\cite{hasenbuschO3,holmO3} As outlined above, the new estimates
will be useful and necessary in connection with the recent fascinating
studies on impurity based questions. In this regard, an extension from
bilayer to planar models has yet to be done.

\textit{Note added.} Recently, a report by Albuquerque
\etal\cite{albuquerque-2008} appeared, which also presents
simulations on the plaquettized Heisenberg model. Since their
motivation is mainly oriented towards showing the applicability of
the contractor renormalization (CORE) method to quantum spin systems,
less emphasis is spent on the analysis of the critical point in
detail.%

\acknowledgments
We acknowledge stimulating discussions with A. Sandvik, S. Wessel,
and A. Muramatsu. We thank L. Bogacz for collaboration in an early
stage of this project. S.W. acknowledges support from the
Studienstiftung des deutsches Volkes, the DFH-UFA under Contract
No. CDFA-02-07, and the Leipzig graduate school ``BuildMoNa.'' This
work was partially performed on the JUMP computer of NIC at the
Forschungszentrum J\"ulich under Project No. HLZ12.

\bibliographystyle{apsrev}   
\bibliography{literature}

\begin{thebibliography}{54}
\expandafter\ifx\csname natexlab\endcsname\relax\def\natexlab#1{#1}\fi
\expandafter\ifx\csname bibnamefont\endcsname\relax
  \def\bibnamefont#1{#1}\fi
\expandafter\ifx\csname bibfnamefont\endcsname\relax
  \def\bibfnamefont#1{#1}\fi
\expandafter\ifx\csname citenamefont\endcsname\relax
  \def\citenamefont#1{#1}\fi
\expandafter\ifx\csname url\endcsname\relax
  \def\url#1{\texttt{#1}}\fi
\expandafter\ifx\csname urlprefix\endcsname\relax\def\urlprefix{URL }\fi
\providecommand{\bibinfo}[2]{#2}
\providecommand{\eprint}[2][]{\url{#2}}

\bibitem[{\citenamefont{Schollw\"ock et~al.}(2004)\citenamefont{Schollw\"ock,
  Richter, Farnell, and Bishop}}]{quantummagnetism-Richter}
\bibinfo{editor}{\bibfnamefont{U.}~\bibnamefont{Schollw\"ock}},
  \bibinfo{editor}{\bibfnamefont{J.}~\bibnamefont{Richter}},
  \bibinfo{editor}{\bibfnamefont{D.}~\bibnamefont{Farnell}}, \bibnamefont{and}
  \bibinfo{editor}{\bibfnamefont{R.}~\bibnamefont{Bishop}} (Eds.),
  Lecture Notes in Physics Vol. 645,   (Springer, Berlin, 2004).

\bibitem[{\citenamefont{Sachdev}(2008)}]{sachdev-2008-4}
\bibinfo{author}{\bibfnamefont{S.}~\bibnamefont{Sachdev}},
  \bibinfo{journal}{Nature Phys.} \textbf{\bibinfo{volume}{4}},
  \bibinfo{pages}{173} (\bibinfo{year}{2008}).

\bibitem[{\citenamefont{Trotzky et~al.}(2008)\citenamefont{Trotzky, Cheinet,
  Folling, Feld, Schnorrberger, Rey, Polkovnikov, Demler, Lukin, and
  Bloch}}]{Trotzky:2008a}
\bibinfo{author}{\bibfnamefont{S.}~\bibnamefont{Trotzky}},
  \bibinfo{author}{\bibfnamefont{P.}~\bibnamefont{Cheinet}},
  \bibinfo{author}{\bibfnamefont{S.}~\bibnamefont{Folling}},
  \bibinfo{author}{\bibfnamefont{M.}~\bibnamefont{Feld}},
  \bibinfo{author}{\bibfnamefont{U.}~\bibnamefont{Schnorrberger}},
  \bibinfo{author}{\bibfnamefont{A.~M.} \bibnamefont{Rey}},
  \bibinfo{author}{\bibfnamefont{A.}~\bibnamefont{Polkovnikov}},
  \bibinfo{author}{\bibfnamefont{E.~A.} \bibnamefont{Demler}},
  \bibinfo{author}{\bibfnamefont{M.~D.} \bibnamefont{Lukin}}, \bibnamefont{and}
  \bibinfo{author}{\bibfnamefont{I.}~\bibnamefont{Bloch}},
  \bibinfo{journal}{Science} \textbf{\bibinfo{volume}{319}},
  \bibinfo{pages}{295} (\bibinfo{year}{2008}).

\bibitem[{\citenamefont{Sachdev}(1999)}]{sachdev:qpt}
\bibinfo{author}{\bibfnamefont{S.}~\bibnamefont{Sachdev}},
  \emph{\bibinfo{title}{Quantum Phase Transitions}} (\bibinfo{publisher}{Cambridge University Press},
  \bibinfo{address}{Cambridge}, \bibinfo{year}{1999}).

\bibitem[{\citenamefont{Vojta}(2003)}]{vojta-2003-66}
\bibinfo{author}{\bibfnamefont{M.}~\bibnamefont{Vojta}},
  \bibinfo{journal}{Rep. Prog. Phys.}
  \textbf{\bibinfo{volume}{66}}, \bibinfo{pages}{2069} (\bibinfo{year}{2003}).

\bibitem[{\citenamefont{Sandvik}(2007)}]{sandvik-2007}
\bibinfo{author}{\bibfnamefont{A.~W.} \bibnamefont{Sandvik}},
  \bibinfo{journal}{Phys. Rev. Lett.} \textbf{\bibinfo{volume}{98}},
  \bibinfo{pages}{227202} (\bibinfo{year}{2007}).

\bibitem[{\citenamefont{Giamarchi et~al.}(2008)\citenamefont{Giamarchi, Ruegg,
  and Tchernyshyov}}]{giamarchi-2008-4}
\bibinfo{author}{\bibfnamefont{T.}~\bibnamefont{Giamarchi}},
  \bibinfo{author}{\bibfnamefont{C.}~\bibnamefont{Ruegg}}, \bibnamefont{and}
  \bibinfo{author}{\bibfnamefont{O.}~\bibnamefont{Tchernyshyov}},
  \bibinfo{journal}{Nature Phys.} \textbf{\bibinfo{volume}{4}},
  \bibinfo{pages}{198} (\bibinfo{year}{2008}).

\bibitem[{\citenamefont{Sandvik and Scalapino}(1994)}]{PhysRevLett.72.2777}
\bibinfo{author}{\bibfnamefont{A.~W.} \bibnamefont{Sandvik}} \bibnamefont{and}
  \bibinfo{author}{\bibfnamefont{D.~J.} \bibnamefont{Scalapino}},
  \bibinfo{journal}{Phys. Rev. Lett.} \textbf{\bibinfo{volume}{72}},
  \bibinfo{pages}{2777} (\bibinfo{year}{1994}).

\bibitem[{\citenamefont{Sandvik et~al.}(1995)\citenamefont{Sandvik, Chubukov,
  and Sachdev}}]{PhysRevB.51.16483}
\bibinfo{author}{\bibfnamefont{A.~W.} \bibnamefont{Sandvik}},
  \bibinfo{author}{\bibfnamefont{A.~V.} \bibnamefont{Chubukov}},
  \bibnamefont{and} \bibinfo{author}{\bibfnamefont{S.}~\bibnamefont{Sachdev}},
  \bibinfo{journal}{Phys. Rev. B} \textbf{\bibinfo{volume}{51}},
  \bibinfo{pages}{16483} (\bibinfo{year}{1995}).

\bibitem[{\citenamefont{Troyer and Sachdev}(1998)}]{PhysRevLett.81.5418}
\bibinfo{author}{\bibfnamefont{M.}~\bibnamefont{Troyer}} \bibnamefont{and}
  \bibinfo{author}{\bibfnamefont{S.}~\bibnamefont{Sachdev}},
  \bibinfo{journal}{Phys. Rev. Lett.} \textbf{\bibinfo{volume}{81}},
  \bibinfo{pages}{5418} (\bibinfo{year}{1998}).

\bibitem[{\citenamefont{Shevchenko et~al.}(2000)\citenamefont{Shevchenko,
  Sandvik, and Sushkov}}]{PhysRevB.61.3475}
\bibinfo{author}{\bibfnamefont{P.~V.} \bibnamefont{Shevchenko}},
  \bibinfo{author}{\bibfnamefont{A.~W.} \bibnamefont{Sandvik}},
  \bibnamefont{and} \bibinfo{author}{\bibfnamefont{O.~P.}
  \bibnamefont{Sushkov}}, \bibinfo{journal}{Phys. Rev. B}
  \textbf{\bibinfo{volume}{61}}, \bibinfo{pages}{3475} (\bibinfo{year}{2000}).

\bibitem[{\citenamefont{Collins and Hamer}(2008)}]{collins:054419}
\bibinfo{author}{\bibfnamefont{A.}~\bibnamefont{Collins}} \bibnamefont{and}
  \bibinfo{author}{\bibfnamefont{C.~J.} \bibnamefont{Hamer}},
  \bibinfo{journal}{Phys. Rev. B} \textbf{\bibinfo{volume}{78}},
  \bibinfo{eid}{054419} (\bibinfo{year}{2008}).

\bibitem[{\citenamefont{Evertz}(2003)}]{evertz-2003-52}
\bibinfo{author}{\bibfnamefont{H.~G.} \bibnamefont{Evertz}},
  \bibinfo{journal}{Adv. Phys.} \textbf{\bibinfo{volume}{52}},
  \bibinfo{pages}{1} (\bibinfo{year}{2003}).

\bibitem[{\citenamefont{Wang et~al.}(2006)\citenamefont{Wang, Beach, and
  Sandvik}}]{wang:014431}
\bibinfo{author}{\bibfnamefont{L.}~\bibnamefont{Wang}},
  \bibinfo{author}{\bibfnamefont{K.~S.~D.} \bibnamefont{Beach}},
  \bibnamefont{and} \bibinfo{author}{\bibfnamefont{A.~W.}
  \bibnamefont{Sandvik}}, \bibinfo{journal}{Phys. Rev. B} \textbf{\bibinfo{volume}{73}}, \bibinfo{eid}{014431}
 (\bibinfo{year}{2006}).

\bibitem[{\citenamefont{H\"{o}glund and Sandvik}(2007)}]{hoeglund:027205}
\bibinfo{author}{\bibfnamefont{K.~H.} \bibnamefont{H\"{o}glund}}
  \bibnamefont{and} \bibinfo{author}{\bibfnamefont{A.~W.}
  \bibnamefont{Sandvik}}, \bibinfo{journal}{Phys. Rev. Lett.}
  \textbf{\bibinfo{volume}{99}}, \bibinfo{eid}{027205}
 (\bibinfo{year}{2007}).

\bibitem[{\citenamefont{Sachdev et~al.}(1999)\citenamefont{Sachdev,
      Buragohain, and Vojta}}]{sachdev-1999-286}
  \bibinfo{author}{\bibfnamefont{S.}~\bibnamefont{Sachdev}},
  \bibinfo{author}{\bibfnamefont{C.}~\bibnamefont{Buragohain}},
  \bibnamefont{and}
  \bibinfo{author}{\bibfnamefont{M.}~\bibnamefont{Vojta}},
  \bibinfo{journal}{Science} \textbf{\bibinfo{volume}{286}},
  \bibinfo{pages}{2479} (\bibinfo{year}{1999}).

\bibitem[{\citenamefont{Anfuso and Eggert}(2006)}]{anfuso-2006-96}
\bibinfo{author}{\bibfnamefont{F.}~\bibnamefont{Anfuso}} \bibnamefont{and}
  \bibinfo{author}{\bibfnamefont{S.}~\bibnamefont{Eggert}},
  \bibinfo{journal}{Phys. Rev. Lett.} \textbf{\bibinfo{volume}{96}},
  \bibinfo{pages}{017204} (\bibinfo{year}{2006}).

\bibitem[{\citenamefont{Sushkov}(2000)}]{sushkov_impurity}
\bibinfo{author}{\bibfnamefont{O.~P.} \bibnamefont{Sushkov}},
  \bibinfo{journal}{Phys. Rev. B} \textbf{\bibinfo{volume}{62}},
  \bibinfo{pages}{12135} (\bibinfo{year}{2000}).

\bibitem[{\citenamefont{Troyer et~al.}(1997)\citenamefont{Troyer, Imada, and
  Ueda}}]{troyer-1997-66}
\bibinfo{author}{\bibfnamefont{M.}~\bibnamefont{Troyer}},
  \bibinfo{author}{\bibfnamefont{M.}~\bibnamefont{Imada}}, \bibnamefont{and}
  \bibinfo{author}{\bibfnamefont{K.}~\bibnamefont{Ueda}}, \bibinfo{journal}{J.
  Phys. Soc. Jpn.} \textbf{\bibinfo{volume}{66}}, \bibinfo{pages}{2957}
  (\bibinfo{year}{1997}).

\bibitem[{\citenamefont{Matsumoto et~al.}(2001)\citenamefont{Matsumoto, Yasuda,
  Todo, and Takayama}}]{PhysRevB.65.014407}
\bibinfo{author}{\bibfnamefont{M.}~\bibnamefont{Matsumoto}},
  \bibinfo{author}{\bibfnamefont{C.}~\bibnamefont{Yasuda}},
  \bibinfo{author}{\bibfnamefont{S.}~\bibnamefont{Todo}}, \bibnamefont{and}
  \bibinfo{author}{\bibfnamefont{H.}~\bibnamefont{Takayama}},
  \bibinfo{journal}{Phys. Rev. B} \textbf{\bibinfo{volume}{65}},
  \bibinfo{pages}{014407} (\bibinfo{year}{2001}).

\bibitem{ChakravartyPRL}
S. Chakravarty, B.~I. Halperin, and D.~R. Nelson, Phys. Rev. Lett. \textbf{60}, 1057 (1988).

\bibitem[{\citenamefont{Chubukov et~al.}(1994)\citenamefont{Chubukov, Sachdev,
  and Ye}}]{chubukov_prb}
\bibinfo{author}{\bibfnamefont{A.~V.} \bibnamefont{Chubukov}},
  \bibinfo{author}{\bibfnamefont{S.}~\bibnamefont{Sachdev}}, \bibnamefont{and}
  \bibinfo{author}{\bibfnamefont{J.}~\bibnamefont{Ye}}, \bibinfo{journal}{Phys.
  Rev. B} \textbf{\bibinfo{volume}{49}}, \bibinfo{pages}{11919}
  (\bibinfo{year}{1994}).


\bibitem[{\citenamefont{Wenzel et~al.}(2008{\natexlab{a}})\citenamefont{Wenzel,
  Bogacz, and Janke}}]{wenzel-JJp-2008}
\bibinfo{author}{\bibfnamefont{S.}~\bibnamefont{Wenzel}},
  \bibinfo{author}{\bibfnamefont{L.}~\bibnamefont{Bogacz}}, \bibnamefont{and}
  \bibinfo{author}{\bibfnamefont{W.}~\bibnamefont{Janke}},
  \bibinfo{journal}{Phys. Rev. Lett.} \textbf{\bibinfo{volume}{101}},
  \bibinfo{pages}{127202} (\bibinfo{year}{2008}{\natexlab{a}}).

\bibitem[{\citenamefont{Ivanov et~al.}(1996)\citenamefont{Ivanov, Kr\"uger, and
  Richter}}]{PhysRevB.53.2633}
\bibinfo{author}{\bibfnamefont{N.~B.} \bibnamefont{Ivanov}},
  \bibinfo{author}{\bibfnamefont{S.~E.} \bibnamefont{Kr\"uger}},
  \bibnamefont{and} \bibinfo{author}{\bibfnamefont{J.}~\bibnamefont{Richter}},
  \bibinfo{journal}{Phys. Rev. B} \textbf{\bibinfo{volume}{53}},
  \bibinfo{pages}{2633} (\bibinfo{year}{1996}).

\bibitem[{\citenamefont{Singh et~al.}(1988)\citenamefont{Singh, Gelfand, and
  Huse}}]{PhysRevLett.61.2484}
\bibinfo{author}{\bibfnamefont{R.~R.~P.} \bibnamefont{Singh}},
  \bibinfo{author}{\bibfnamefont{M.~P.} \bibnamefont{Gelfand}},
  \bibnamefont{and} \bibinfo{author}{\bibfnamefont{D.~A.} \bibnamefont{Huse}},
  \bibinfo{journal}{Phys. Rev. Lett.} \textbf{\bibinfo{volume}{61}},
  \bibinfo{pages}{2484} (\bibinfo{year}{1988}).

\bibitem[{\citenamefont{Kr\"uger et~al.}(2000)\citenamefont{Kr\"uger, Richter,
  Schulenburg, Farnell, and Bishop}}]{PhysRevB.61.14607}
\bibinfo{author}{\bibfnamefont{S.~E.} \bibnamefont{Kr\"uger}},
  \bibinfo{author}{\bibfnamefont{J.}~\bibnamefont{Richter}},
  \bibinfo{author}{\bibfnamefont{J.}~\bibnamefont{Schulenburg}},
  \bibinfo{author}{\bibfnamefont{D.~J.~J.} \bibnamefont{Farnell}},
  \bibnamefont{and} \bibinfo{author}{\bibfnamefont{R.~F.}
  \bibnamefont{Bishop}}, \bibinfo{journal}{Phys. Rev. B}
  \textbf{\bibinfo{volume}{61}}, \bibinfo{pages}{14607} (\bibinfo{year}{2000}).

\bibitem[{\citenamefont{Tomczak and Richter}(2001)}]{tomczak}
\bibinfo{author}{\bibfnamefont{P.}~\bibnamefont{Tomczak}} \bibnamefont{and}
  \bibinfo{author}{\bibfnamefont{J.}~\bibnamefont{Richter}},
  \bibinfo{journal}{J. Phys. A: Math. Gen.} \textbf{\bibinfo{volume}{34}},
  \bibinfo{pages}{L461} (\bibinfo{year}{2001}).

\bibitem[{\citenamefont{Darradi et~al.}(2004)\citenamefont{Darradi, Richter,
  and Krueger}}]{darradi-2004-16}
\bibinfo{author}{\bibfnamefont{R.}~\bibnamefont{Darradi}},
  \bibinfo{author}{\bibfnamefont{J.}~\bibnamefont{Richter}}, \bibnamefont{and}
  \bibinfo{author}{\bibfnamefont{S.~E.} \bibnamefont{Kr\"uger}},
  \bibinfo{journal}{Cond. Matt.} \textbf{\bibinfo{volume}{16}},
  \bibinfo{pages}{2681} (\bibinfo{year}{2004}).

\bibitem[{\citenamefont{L\"{o}w}(2007)}]{loewNeel}
\bibinfo{author}{\bibfnamefont{U.}~\bibnamefont{L\"{o}w}},
  \bibinfo{journal}{Phys. Rev. B} \textbf{\bibinfo{volume}{76}}, 220409(R) (\bibinfo{year}{2007}).

\bibitem[{\citenamefont{Katoh and Imada}(1993)}]{katoh1993}
\bibinfo{author}{\bibfnamefont{N.}~\bibnamefont{Katoh}} \bibnamefont{and}
  \bibinfo{author}{\bibfnamefont{M.}~\bibnamefont{Imada}}, \bibinfo{journal}{J.
  Phys. Soc. Jpn.} \textbf{\bibinfo{volume}{62}}, \bibinfo{pages}{3728}
  (\bibinfo{year}{1993}).

\bibitem[{\citenamefont{Nohadani et~al.}(2005)\citenamefont{Nohadani, Wessel,
  and Haas}}]{nohadani-2005-72}
\bibinfo{author}{\bibfnamefont{O.}~\bibnamefont{Nohadani}},
  \bibinfo{author}{\bibfnamefont{S.}~\bibnamefont{Wessel}}, \bibnamefont{and}
  \bibinfo{author}{\bibfnamefont{S.}~\bibnamefont{Haas}},
  \bibinfo{journal}{Phys. Rev. B} \textbf{\bibinfo{volume}{72}},
  \bibinfo{pages}{024440} (\bibinfo{year}{2005}).

\bibitem[{\citenamefont{Nohadani et~al.}(2004)\citenamefont{Nohadani, Wessel,
  Normand, and Haas}}]{nohadani-2004-69}
\bibinfo{author}{\bibfnamefont{O.}~\bibnamefont{Nohadani}},
  \bibinfo{author}{\bibfnamefont{S.}~\bibnamefont{Wessel}},
  \bibinfo{author}{\bibfnamefont{B.}~\bibnamefont{Normand}}, \bibnamefont{and}
  \bibinfo{author}{\bibfnamefont{S.}~\bibnamefont{Haas}},
  \bibinfo{journal}{Phys. Rev. B} \textbf{\bibinfo{volume}{69}},
  \bibinfo{pages}{220402(R)} (\bibinfo{year}{2004}).


\bibitem[{\citenamefont{Yasuda et~al.}(2001)\citenamefont{Yasuda, Todo,
  Matsumoto, and Takayama}}]{chitoshi_dilution}
\bibinfo{author}{\bibfnamefont{C.}~\bibnamefont{Yasuda}},
  \bibinfo{author}{\bibfnamefont{S.}~\bibnamefont{Todo}},
  \bibinfo{author}{\bibfnamefont{M.}~\bibnamefont{Matsumoto}},
  \bibnamefont{and} \bibinfo{author}{\bibfnamefont{H.}~\bibnamefont{Takayama}},
  \bibinfo{journal}{Phys. Rev. B} \textbf{\bibinfo{volume}{64}},
  \bibinfo{pages}{092405} (\bibinfo{year}{2001}).

\bibitem[{\citenamefont{Senthil et~al.}(2005)\citenamefont{Senthil, Balents,
  Sachdev, Vishwanath, and Fisher}}]{senthil-2005-74}
\bibinfo{author}{\bibfnamefont{T.}~\bibnamefont{Senthil}},
  \bibinfo{author}{\bibfnamefont{L.}~\bibnamefont{Balents}},
  \bibinfo{author}{\bibfnamefont{S.}~\bibnamefont{Sachdev}},
  \bibinfo{author}{\bibfnamefont{A.}~\bibnamefont{Vishwanath}},
  \bibnamefont{and} \bibinfo{author}{\bibfnamefont{M.~P.~A.}
  \bibnamefont{Fisher}}, \bibinfo{journal}{J. Phys. Soc. Jpn. Suppl.} \textbf{\bibinfo{volume}{74}}, \bibinfo{pages}{1}
  (\bibinfo{year}{2005}).

\bibitem[{\citenamefont{Koga et~al.}(1999)\citenamefont{Koga, Kumada, and
  Kawakami}}]{koga}
\bibinfo{author}{\bibfnamefont{A.}~\bibnamefont{Koga}},
  \bibinfo{author}{\bibfnamefont{S.}~\bibnamefont{Kumada}}, \bibnamefont{and}
  \bibinfo{author}{\bibfnamefont{N.}~\bibnamefont{Kawakami}},
  \bibinfo{journal}{J. Soc. Phy. Jpn} \textbf{\bibinfo{volume}{68}},
  \bibinfo{pages}{642} (\bibinfo{year}{1999}).

\bibitem[{\citenamefont{Sirker et~al.}(2002)\citenamefont{Sirker, Kl\"{u}mper,
  and Hamacher}}]{sirker:134409}
\bibinfo{author}{\bibfnamefont{J.}~\bibnamefont{Sirker}},
  \bibinfo{author}{\bibfnamefont{A.}~\bibnamefont{Kl\"{u}mper}},
  \bibnamefont{and} \bibinfo{author}{\bibfnamefont{K.}~\bibnamefont{Hamacher}},
  \bibinfo{journal}{Phys. Rev. B} \textbf{\bibinfo{volume}{65}},
  \bibinfo{pages}{134409} (\bibinfo{year}{2002}).

\bibitem[{\citenamefont{Singh et~al.}(1999)\citenamefont{Singh, Weihong, Hamer,
  and Oitmaa}}]{singh-1999-60}
\bibinfo{author}{\bibfnamefont{R.~R.~P.} \bibnamefont{Singh}},
  \bibinfo{author}{\bibfnamefont{Z.}~\bibnamefont{Weihong}},
  \bibinfo{author}{\bibfnamefont{C.~J.} \bibnamefont{Hamer}}, \bibnamefont{and}
  \bibinfo{author}{\bibfnamefont{J.}~\bibnamefont{Oitmaa}},
  \bibinfo{journal}{Phys. Rev. B} \textbf{\bibinfo{volume}{60}},
  \bibinfo{pages}{7278} (\bibinfo{year}{1999}).

\bibitem[{\citenamefont{L\"auchli et~al.}(2002)\citenamefont{L\"auchli, Wessel,
  and Sigrist}}]{wessel:shastry}
\bibinfo{author}{\bibfnamefont{A.}~\bibnamefont{L\"auchli}},
  \bibinfo{author}{\bibfnamefont{S.}~\bibnamefont{Wessel}}, \bibnamefont{and}
  \bibinfo{author}{\bibfnamefont{M.}~\bibnamefont{Sigrist}},
  \bibinfo{journal}{Phys. Rev. B} \textbf{\bibinfo{volume}{66}},
  \bibinfo{pages}{014401} (\bibinfo{year}{2002}).

\bibitem[{\citenamefont{Capponi et~al.}(2004)\citenamefont{Capponi,
  L\"{a}uchli, and Mambrini}}]{capponi:104424}
\bibinfo{author}{\bibfnamefont{S.}~\bibnamefont{Capponi}},
  \bibinfo{author}{\bibfnamefont{A.}~\bibnamefont{L\"{a}uchli}},
  \bibnamefont{and} \bibinfo{author}{\bibfnamefont{M.}~\bibnamefont{Mambrini}},
  \bibinfo{journal}{Phys. Rev. B} \textbf{\bibinfo{volume}{70}},
  \bibinfo{pages}{104424} (\bibinfo{year}{2004}).

\bibitem[{\citenamefont{Sandvik and Kurkij\"arvi}(1991)}]{PhysRevB.43.5950}
\bibinfo{author}{\bibfnamefont{A.~W.} \bibnamefont{Sandvik}} \bibnamefont{and}
  \bibinfo{author}{\bibfnamefont{J.}~\bibnamefont{Kurkij\"arvi}},
  \bibinfo{journal}{Phys. Rev. B} \textbf{\bibinfo{volume}{43}},
  \bibinfo{pages}{5950} (\bibinfo{year}{1991}).

\bibitem[{\citenamefont{Sandvik}(1999)}]{PhysRevB.59.R14157}
\bibinfo{author}{\bibfnamefont{A.~W.} \bibnamefont{Sandvik}},
  \bibinfo{journal}{Phys. Rev. B} \textbf{\bibinfo{volume}{59}},
  \bibinfo{pages}{R14157} (\bibinfo{year}{1999}).

\bibitem[{\citenamefont{Sylju\aa{}sen and Sandvik}(2002)}]{PhysRevE.66.046701}
\bibinfo{author}{\bibfnamefont{O.~F.} \bibnamefont{Sylju\aa{}sen}}
  \bibnamefont{and} \bibinfo{author}{\bibfnamefont{A.~W.}
  \bibnamefont{Sandvik}}, \bibinfo{journal}{Phys. Rev. E}
  \textbf{\bibinfo{volume}{66}}, \bibinfo{pages}{046701}
  (\bibinfo{year}{2002}).

\bibitem[{\citenamefont{Alet et~al.}(2005)\citenamefont{Alet, Wessel, and
  Troyer}}]{alet:036706}
\bibinfo{author}{\bibfnamefont{F.}~\bibnamefont{Alet}},
  \bibinfo{author}{\bibfnamefont{S.}~\bibnamefont{Wessel}}, \bibnamefont{and}
  \bibinfo{author}{\bibfnamefont{M.}~\bibnamefont{Troyer}},
  \bibinfo{journal}{Phys. Rev. E} \textbf{\bibinfo{volume}{71}}, \bibinfo{eid}{036706}  (\bibinfo{year}{2005}).

\bibitem[{\citenamefont{Wang and Landau}(2001)}]{PhysRevLett.86.2050}
\bibinfo{author}{\bibfnamefont{F.}~\bibnamefont{Wang}} \bibnamefont{and}
  \bibinfo{author}{\bibfnamefont{D.~P.} \bibnamefont{Landau}},
  \bibinfo{journal}{Phys. Rev. Lett.} \textbf{\bibinfo{volume}{86}},
  \bibinfo{pages}{2050} (\bibinfo{year}{2001}).

\bibitem[{\citenamefont{Troyer et~al.}(2003)\citenamefont{Troyer, Wessel, and
  Alet}}]{PhysRevLett.90.120201}
\bibinfo{author}{\bibfnamefont{M.}~\bibnamefont{Troyer}},
  \bibinfo{author}{\bibfnamefont{S.}~\bibnamefont{Wessel}}, \bibnamefont{and}
  \bibinfo{author}{\bibfnamefont{F.}~\bibnamefont{Alet}},
  \bibinfo{journal}{Phys. Rev. Lett.} \textbf{\bibinfo{volume}{90}},
  \bibinfo{pages}{120201} (\bibinfo{year}{2003}).

\bibitem[{\citenamefont{{A. M. Ferrenberg} and {R.
  Swendsen}}(1989)}]{ferrenberg:multi}
\bibinfo{author}{\bibnamefont{{A. M. Ferrenberg}}} \bibnamefont{and}
  \bibinfo{author}{\bibnamefont{{R. H. Swendsen}}}, \bibinfo{journal}{Phys. Rev.
  Lett.} \textbf{\bibinfo{volume}{63}}, \bibinfo{pages}{1195} (\bibinfo{year}{1989}).

\bibitem[{\citenamefont{Troyer et~al.}(2004)\citenamefont{Troyer, Wessel, and
  Alet}}]{Troyer_multihist}
\bibinfo{author}{\bibfnamefont{M.}~\bibnamefont{Troyer}},
  \bibinfo{author}{\bibfnamefont{F.}~\bibnamefont{Alet}}, \bibnamefont{and}
  \bibinfo{author}{\bibfnamefont{S.}~\bibnamefont{Wessel}},
  \bibinfo{journal}{Braz. J. of Phys.} \textbf{\bibinfo{volume}{34}},
  \bibinfo{pages}{377} (\bibinfo{year}{2004}).

\bibitem[{\citenamefont{Sandvik}(1997)}]{PhysRevB.56.11678}
\bibinfo{author}{\bibfnamefont{A.~W.} \bibnamefont{Sandvik}},
  \bibinfo{journal}{Phys. Rev. B} \textbf{\bibinfo{volume}{56}},
  \bibinfo{pages}{11678} (\bibinfo{year}{1997}).

\bibitem[{\citenamefont{Fisher et~al.}(1989)\citenamefont{Fisher, Weichman,
  Grinstein, and Fisher}}]{PhysRevB.40.546}
\bibinfo{author}{\bibfnamefont{M.~P.~A.} \bibnamefont{Fisher}},
  \bibinfo{author}{\bibfnamefont{P.~B.} \bibnamefont{Weichman}},
  \bibinfo{author}{\bibfnamefont{G.}~\bibnamefont{Grinstein}},
  \bibnamefont{and} \bibinfo{author}{\bibfnamefont{D.~S.}
  \bibnamefont{Fisher}}, \bibinfo{journal}{Phys. Rev. B}
  \textbf{\bibinfo{volume}{40}}, \bibinfo{pages}{546} (\bibinfo{year}{1989}).

\bibitem[{\citenamefont{Beach et~al.}(2005)\citenamefont{Beach, Wang, and
  Sandvik}}]{beach-2005}
\bibinfo{author}{\bibfnamefont{K.~S.~D.} \bibnamefont{Beach}},
  \bibinfo{author}{\bibfnamefont{L.}~\bibnamefont{Wang}}, \bibnamefont{and}
  \bibinfo{author}{\bibfnamefont{A.~W.} \bibnamefont{Sandvik}},
  arXiv:0505194 (unpublished) (\bibinfo{year}{2005}).

\bibitem[{\citenamefont{Wenzel et~al.}(2008{\natexlab{b}})\citenamefont{Wenzel,
  Bittner, Janke, and Schakel}}]{wenzel-2007}
\bibinfo{author}{\bibfnamefont{S.}~\bibnamefont{Wenzel}},
  \bibinfo{author}{\bibfnamefont{E.}~\bibnamefont{Bittner}},
  \bibinfo{author}{\bibfnamefont{W.}~\bibnamefont{Janke}}, \bibnamefont{and}
  \bibinfo{author}{\bibfnamefont{A.~M.~J.} \bibnamefont{Schakel}},
  \bibinfo{journal}{Nucl. Phys. B} \textbf{\bibinfo{volume}{793}},
  \bibinfo{pages}{344} (\bibinfo{year}{2008}{\natexlab{b}}).

\bibitem[{\citenamefont{Binder}(1981)}]{Binder1981}
\bibinfo{author}{\bibfnamefont{K.}~\bibnamefont{Binder}}, \bibinfo{journal}{Z.
  Phys. B} \textbf{\bibinfo{volume}{43}}, 119 (\bibinfo{year}{1981}).

\bibitem[{\citenamefont{Efron}(1982)}]{efron}
\bibinfo{author}{\bibfnamefont{B.}~\bibnamefont{Efron}},
  \emph{\bibinfo{title}{The Jackknife, the Bootstrap, and other Resampling
  Plans}} (\bibinfo{publisher}{Society for Industrial and Applied Mathematics
  [SIAM]}, \bibinfo{address}{Philadelphia}, \bibinfo{year}{1982}).

\bibitem[{\citenamefont{Campostrini et~al.}(2002)\citenamefont{Campostrini,
  Hasenbusch, Pelissetto, Rossi, and Vicari}}]{hasenbuschO3}
\bibinfo{author}{\bibfnamefont{M.}~\bibnamefont{Campostrini}},
  \bibinfo{author}{\bibfnamefont{M.}~\bibnamefont{Hasenbusch}},
  \bibinfo{author}{\bibfnamefont{A.}~\bibnamefont{Pelissetto}},
  \bibinfo{author}{\bibfnamefont{P.}~\bibnamefont{Rossi}}, \bibnamefont{and}
  \bibinfo{author}{\bibfnamefont{E.}~\bibnamefont{Vicari}},
  \bibinfo{journal}{Phys. Rev. B} \textbf{\bibinfo{volume}{65}},
  \bibinfo{pages}{144520} (\bibinfo{year}{2002}).

\bibitem[{\citenamefont{Holm and Janke}(1993)}]{holmO3}
\bibinfo{author}{\bibfnamefont{C.}~\bibnamefont{Holm}} \bibnamefont{and}
  \bibinfo{author}{\bibfnamefont{W.}~\bibnamefont{Janke}},
  \bibinfo{journal}{Phys. Rev. B} \textbf{\bibinfo{volume}{48}},
  \bibinfo{pages}{936} (\bibinfo{year}{1993}).

\bibitem[{\citenamefont{Albuquerque et~al.}(2008)\citenamefont{Albuquerque,
  Troyer, and Oitmaa}}]{albuquerque-2008}
\bibinfo{author}{\bibfnamefont{A.~F.} \bibnamefont{Albuquerque}},
  \bibinfo{author}{\bibfnamefont{M.}~\bibnamefont{Troyer}}, \bibnamefont{and}
  \bibinfo{author}{\bibfnamefont{J.}~\bibnamefont{Oitmaa}},
Phys. Rev. B \textbf{78}, 132402 (2008).

\end{thebibliography}

\end{document}